\documentclass[1p]{elsarticle}

\usepackage{lineno,hyperref}
\usepackage[english]{babel}
\usepackage{subfig}
\usepackage{multirow}
\usepackage{amsmath}

\usepackage{verbatim} %% this one's only to use the \begin{comment} stuff so can be removed lateribl

\journal{Nuclear Physics A}

\begin{document}

%\linenumbers

\begin{frontmatter}

\title{Search for the double-beta decay of $^{82}$Se to the excited states of $^{82}$Kr with NEMO-3}
%% Author List

\author[IPHC]{R.~Arnold}
\author[LAL]{C.~Augier}
%\author[INL]{J.D.~Baker\fnref{fnRef}}
\author[ITEP]{A.S.~Barabash}
\author[UCL]{A.~Basharina-Freshville}
\author[LAL]{S.~Blondel}
\author[Man]{S.~Blot}
\author[LAL]{M.~Bongrand}
\author[LAL]{D.~Boursette}
\author[Com]{R.~Breier}
\author[JINR,NRNUM]{V.~Brudanin}
% ...
%\\
\vspace{1cm}
\author[]{The NEMO-3 collaboration}

\address[IPHC]{IPHC, ULP, CNRS/IN2P3, F-67037 Strasbourg, France}
\address[LAL]{LAL, Univ Paris-Sud, CNRS/IN2P3, F-91405 Orsay, France}
\address[ITEP]{NRC "Kurchatov Institute" - ITEP, 117218 Moscow, Russia}
\address[UCL]{UCL, London WC1E 6BT, United Kingdom}
\address[Man]{University of Manchester, Manchester M13 9PL,~United Kingdom}
\address[JINR]{JINR, 141980 Dubna, Russia}
\address[NRNUM]{National Research Nuclear University MEPhI, 115409 Moscow, Russia}
% ...
\address[Com]{FMFI,~Comenius~Univ.,~SK-842~48~Bratislava,~Slovakia}
% ...

%\collaboration{NEMO-3 Collaboration}

\cortext[CorrRef]{Corresponding authors}
\fntext[fnRef]{Deceased}

%\begin{center}
%{\it (The NEMO-3 collaboration)}
%\end{center}

\begin{abstract}
%\begin{linenumbers}
The double-beta decay of $^{82}$Se to the 0$_{1}^{+}$ excited state
of $^{82}$Kr has been studied with the NEMO-3 detector using 0.93 kg
of enriched $^{82}$Se measured for 4.75 y, corresponding to an
exposure of 4.42 kg$^.$y. A dedicated analysis to reconstruct the
$\gamma$-rays has been performed to search for events in the 2e2$\gamma$
channel. No evidence of a $2\nu\beta\beta$ decay to the 0$_{1}^{+}$
state has been observed and a limit  of
$T_{1/2}^{2\nu}({}^{\mathrm{82}}\mathrm{Se},0_{gs}^{+}\rightarrow0_{1}^{+})>1.3\times10^{21}\:\mathrm{y}$
at 90$\%$~CL has been set. Concerning the $0\nu\beta\beta$ decay to
the 0$_{1}^{+}$ state, a limit for this decay has been obtained
with
$T_{1/2}^{0\nu}({}^{\mathrm{82}}\mathrm{Se},0_{gs}^{+}\rightarrow0_{1}^{+})>2.3\times10^{22}\:\mathrm{y}$
at 90$\%$~CL, independently from the $2\nu\beta\beta$ decay
process. These results are obtained for the first
time with a
tracko-calo detector, reconstructing every particle in the final state.
%\end{linenumbers}
\end{abstract}


\end{frontmatter}

\section{Introduction}
\label{sec:intro}
The search for the neutrinoless double-beta decay 
($0\nu\beta\beta$) is of major importance in neutrino and particle
physics. Its observation would prove the Majorana nature of the
neutrino and would be the first evidence for lepton number
violation. Up to now, no evidence of such a process has been found and
the best half-life limits are in the $10^{24}$-$10^{26}$ y range
\cite{NEMO32015, CUORE2018, GERDA2018, Gando2016}.    
%\cite{NEMO32015}\cite{CUORE2018}\cite{GERDA2018}\cite{Gando2016}.    
$^{82}$Se is one of the best isotopes to investigate
$0\nu\beta\beta$ decay. In particular, its high Q$_{\beta\beta}$-value
of 2997.9$\pm$0.3 keV \cite{Lincoln2013} lies above the main
backgrounds coming from natural radioactivity. There exist also
well-known methods of Se isotopic enrichment through
centrifugal separation. This is why $^{82}$Se is the baseline isotope for
past, current or future experiments such as LUCIFER \cite{LUCIFER}, CUPID-0
\cite{CUPID-0_v2-detector} and
SuperNEMO \cite{SuperNEMO}. Several studies have been performed in the
past to search for $0\nu\beta\beta$ decay of $^{82}$Se to the ground state of
$^{82}$Kr and recently new limits on the half-life have been obtained
with the NEMO-3 (2.5$\times 10^{23}$ y
\cite{Arnold:2018se}) and CUPID-0 experiments (3.5$\times 10^{24}$ y
\cite{CUPID-0_v2}).

The double-beta decay with emission of two neutrinos ($2\nu\beta\beta$) is a second order electroweak process in the Standard Model. It allows the experimental determination of the Nuclear Matrix Elements (NME) for such processes and provides a robust test for the different nuclear models. It could constrain the quenching factor of the axial-vector coupling constant g$_A$ and give the possibility to improve the quality of NME calculations for $0\nu\beta\beta$ decay \cite{MENENDEZ2009139,Barea2015,Hyvarinen2016,Dolinski2019}.
This process has been observed for 11 double-beta isotopes with a
range of measured half-lives between $10^{18}$-$10^{24}$ y
\cite{Barabash2015,Barabash2019}. For
$^{82}$Se, several experiments have measured the $2\nu\beta\beta$ decay to the ground state with the most precise half-life value to date of $\mathrm{9.39\pm0.17(stat)\pm0.58(syst)\times10^{19}}$ y measured with the NEMO-3 experiment \cite{Arnold:2018se}. 

The search for $\beta\beta$ decay to excited states is also an
interesting way to study such processes. Indeed, these decays have a
very clear-cut signature using the 2e1$\gamma$ channel (to the
2$_{1}^{+}$ state) or the 2e2$\gamma$ channel (to the 0$_{1}^{+}$ or
2$_{2}^{+}$ state) which can dramatically reduce the number of background
events. The disadvantages are a lower Q$_{\beta\beta}$ available energy
for electrons which suppresses the probability of the decay and a
lower detection efficiency for electrons and $\gamma$-rays. Nevertheless, the
decay to excited states is of importance to test the nuclear matrix
elements. A detailed analysis for $2 \nu \beta \beta$ decay of $^{100}$Mo and $^{150}$Nd to
the excited $0^+_1$ state of $^{100}$Ru and $^{150}$Sm, respectively,
showed that corresponding NME are only suppressed by $\sim$ 30$\%$ when
compared with the NME to ground state transition \cite{NEMO3:FirstResult,MoExciteHPGe,excitedMo,NEMO3:150Nd,NdExciteHPGe,NdExciteHPGe2,Iachello2012}.
%[15-21].
%\cite{NEMO3:FirstResult}\cite{MoExciteHPGe}\cite{excitedMo}\cite{NEMO3:150Nd}\cite{NdExciteHPGe}\cite{NdExciteHPGe2}\cite{Iachello2012}.
  
Up to now, the $2\nu\beta\beta$ decay to excited states has only been
observed for two isotopes: $^{100}$Mo and $^{150}$Nd with typical
half-lives of 10$^{20}$-10$^{21}$ y \cite{ReviewBarabash}. It is important to note that
this decay has been observed only to the 0$_{1}^{+}$ excited state
(with the emission of two $\gamma$-rays) which is favoured compared to the
decay to the 2$_{1}^{+}$ or 2$_{2}^{+}$ excited states. These measurements have been performed
using both High Purity Germanium (HPGe) detectors by measuring only
the $\gamma$-rays in the cascade
%\cite{MoExciteHPGe}\cite{NdExciteHPGe}\cite{NdExciteHPGe2}[23-27] and
\cite{Barabash1995,Barabash1999,Braeckeleer,NdExciteHPGe,Kidd2009,Belli,MoExciteHPGe,NdExciteHPGe2}
and
%[16-23] and
``tracker-calorimeter'' detectors such as NEMO-3 able to measure the
energies of both electrons and $\gamma$-rays
\cite{excitedMo,theseSophie}. For $^{82}$Se, there is up to now
no evidence for such a decay. Stringent limits have been
obtained by the LUCIFER collaboration for the (2$\nu$+0$\nu$)$\beta\beta$
decay to various excited states of $^{82}$Kr using a HPGe detector
\cite{LUCIFER-results}. Nevertheless, this technique using only $\gamma$-rays does not allow to distinguish between $2\nu\beta\beta$ and $0\nu\beta\beta$.  More recently, more stringent limits
have been set by the CUPID-0 collaboration for the 0$\nu\beta\beta$
decay to various excited states of $^{82}$Kr using ZnSe 
scintillating bolometers \cite{CUPID-0}. 

\begin{figure}[htpb]
  \centering
  \includegraphics[width=0.6\textwidth]{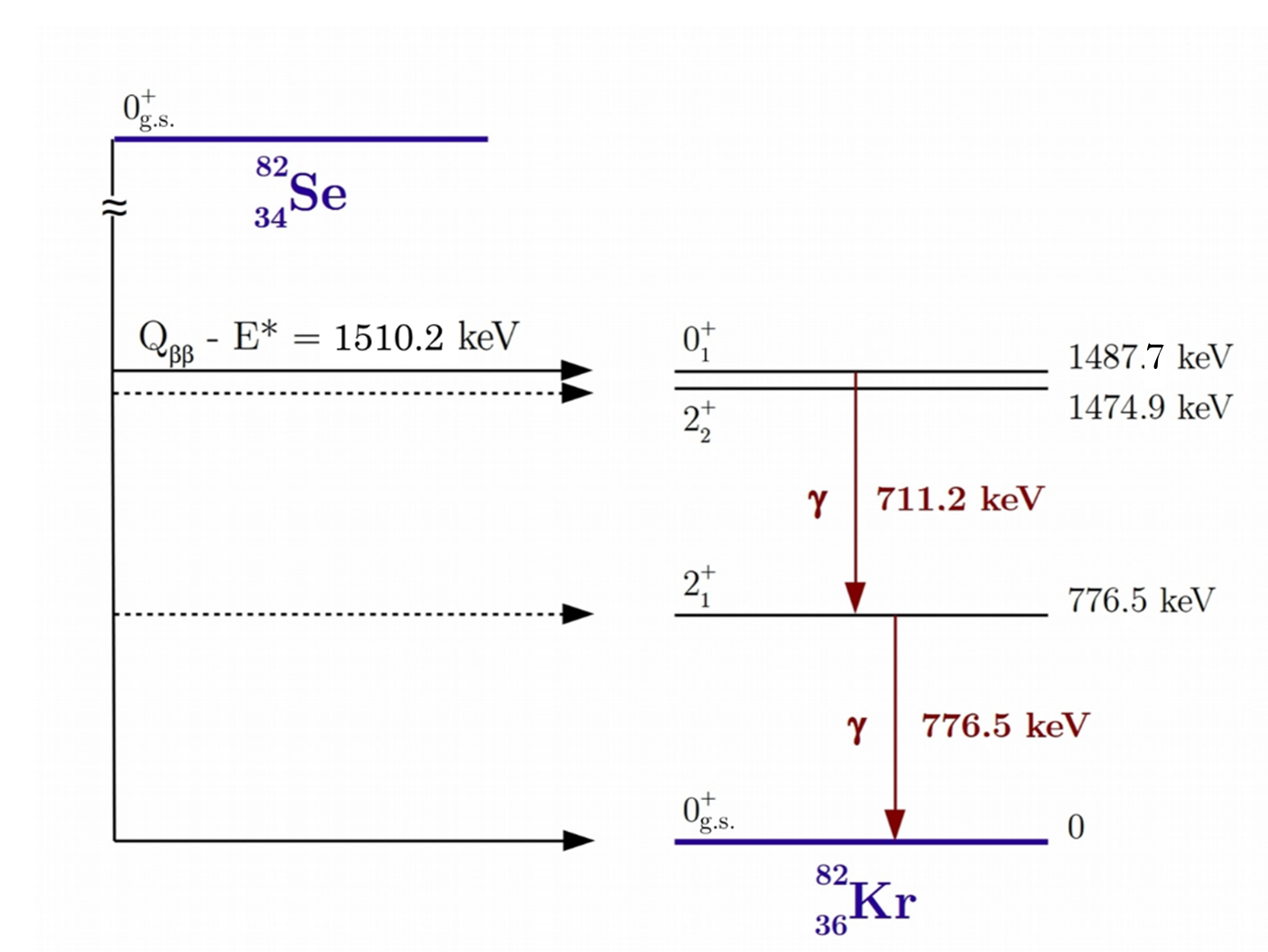}
  \caption{Decay scheme of the $^{82}$Se $\beta\beta$ decay to the
    0$_{1}^{+}$ excited state with the emission of two electrons
    sharing 1510.2 keV and two prompt $\gamma$-rays with energies of 711.2 and 776.5
    keV \cite{TableOfIsotopes}.} 
  \label{fig:se82-decay-fig}
\end{figure}

In this work, we will present a detailed study of the $^{82}$Se
$2\nu\beta\beta$ and $0\nu\beta\beta$ decays to the 0$_{1}^{+}$
excited state of $^{82}$Kr, expected to be the most favoured \cite{QRPAWS,QRPA2nu},
with the full exposure of the NEMO-3 experiment. In this analysis, we
have access to the full topology of the decay. It consists of the
emission of two electrons sharing 1510.2 keV of energy and accompanied by two $\gamma$-rays with energies of 711.2 keV and 776.5 keV respectively, as illustrated in Figure \ref{fig:se82-decay-fig}. After a presentation of the NEMO-3 detector, the $^{82}$Se source foils and the associated backgrounds, we will present a dedicated analysis tool called gamma tracking (GT) developed to reconstruct efficiently the $\gamma$-rays in such a decay. Finally, we will present the results of the $2\nu\beta\beta$ and $0\nu\beta\beta$ decays of $^{82}$Se to the $0_{1}^{+}$ excited state of $^{82}$Kr with the full NEMO-3 exposure of 4.42 kg$\cdot$y.

\section{\texorpdfstring{NEMO-3 detector, $^{82}$Se source foils and associated backgrounds}{NEMO-3}}

\subsection{NEMO-3 detector}

\begin{figure}[htpb]
  \centering
  \includegraphics[width=0.6\textwidth]{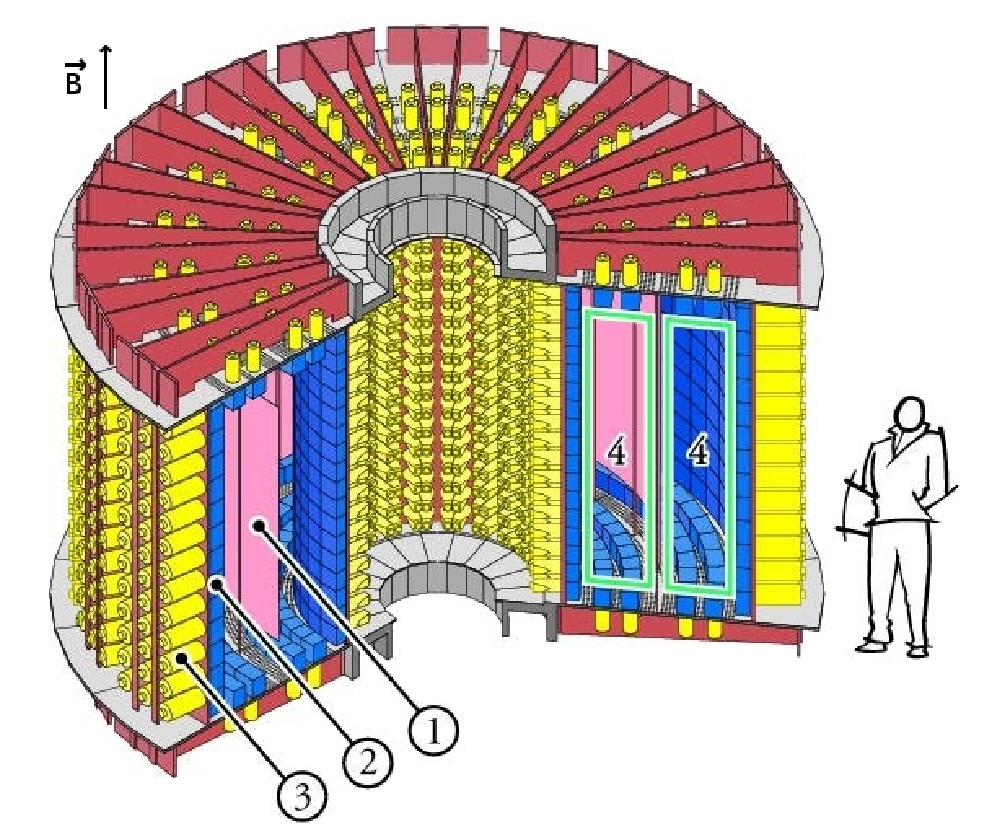}
  \caption{Cross-sectional view of the NEMO-3 detector. The detector
    consists of source foils (1), scintillators (2), photomultipliers (3) and a wire chamber (4).} 
  \label{fig:NEMO3-fig}
\end{figure}

NEMO-3 was a detector installed in the Modane Underground Laboratory
(LSM) under 4800 m water-equivalent in order to be protected against cosmic muons. It took data from February 2003 to January 2011. It consisted of a hollow cylinder divided into 20 sectors hosting thin source foils from 7 different enriched isotopes with a typical thickness of approximately 50 mg/cm$^2$ (as shown in Figure \ref{fig:NEMO3-fig}). The main isotope to search for $0\nu\beta\beta$ decay was $^{100}$Mo with a total mass of 6.914 kg. The second isotope of interest was $^{82}$Se with a mass of 0.932 kg shared in 3 sectors. The five other isotopes studied were by decreasing order of mass: $^{130}$Te (0.454 kg),$^{116}$Cd
(0.405 kg),  $^{150}$Nd (36.55 g), $^{96}$Zr
(9.43 g), and $^{48}$Ca (6.99~g) (see \cite{NEMO32015,Arnold:2004xq} for more details).
 
The source foils were hung at the center of a wire chamber composed
of 6180 cells operating in Geiger mode. The gas was a mixture composed
of 94.85\% helium, 4\% ethanol, 1\% argon and 0.15\% of water
vapour. These cells were placed inside a 25 G magnetic field produced
by a solenoid surrounding the detector. Charged particles thus had a
curved trajectory when crossing the tracking chamber, which allowed the
identification of a negative curvature for 95\% of electrons at 1
MeV. The minimal distance traveled by a particle crossing the tracker
is $\sim$1.1 m, which corresponds to a typical minimal time of flight of ~ 3 ns.
The resolution of the tracker was 0.5 mm transverse to the wires
and 8 mm in the vertical direction for 1 MeV electrons.

A calorimeter enclosed the wire chamber. It was made from 1940 plastic
scintillator blocks, each one with a typical size of 20 x 20 x 10
$\mathrm{cm^3}$ and coupled to a low background
photomultiplier (PMT) through a light guide. The calorimeter measured
the kinetic energy of the particles and the time difference between
two distant hits could be recorded. The blocks had an energy resolution
of $6-7\% /\sqrt \mathrm{E(MeV)} $  and a time resolution of 250 ps ($\sigma$ at 1 MeV).

NEMO-3 was a unique detector as it combined tracking and calorimetry techniques. A charged particle (e$^{-}$, e$^{+}$...) was identified when going across and ionizing the wire chamber gas. Its track was associated to an energy deposit in a calorimeter block neighbouring the last fired Geiger cell. $\gamma$-rays were identified when energy was deposited in a calorimeter block but no track was associated. Alpha particles were identified as straight, short tracks as they could not travel more than $\sim 40$ cm in the tracker due to their high ionisation energy loss.

In order to run in low background conditions, the NEMO-3 detector had
to be protected from natural radioactivity. To do so, a passive
shielding of 19 cm iron was surrounding the detector in order to stop external $\gamma$-rays. In addition, borated water, paraffin and wood were also used to moderate and absorb the environmental neutrons. For a more detailed description of the NEMO-3 detector, see \cite{Arnold:2004xq}.

\subsection{$^{82}$Se source}
Two different batches of $^{82}$Se source were used (referred to as $^{82}$Se(I) and $^{82}$Se(II)). Those batches had an enrichment factor of $97.02 \pm 0.05 \%$ and $96.82 \pm 0.05 \%$ respectively.  To produce source foils, the enriched  $^{82}$Se powder was mixed with 
polyvinyl alcohol (PVA) glue and
deposited between $\sim$23 $\mu$m thick Mylar foils, producing composite source foils. The total mass of the $^{82}$Se isotope in NEMO-3 was $932.4 \pm 5.0 ~\mbox{g}$. An analysis of these $^{82}$Se foils was conducted in order to search for $2\nu\beta\beta$ and $0\nu\beta\beta$ decays to ground state and is detailed in \cite{Arnold:2018se}.

\subsection{Backgrounds}
\label{sec:bckg}

 With its powerful topology reconstruction ability, the NEMO-3 detector was able to identify 2e2$\gamma$ events that were selected for $\beta\beta$ decay to excited states. However, some background isotopes could also produce this type of event. Among them, $\mathrm{^{214}Bi}$ and $\mathrm{^{208}Tl}$ decays were the main sources of background as the produced particles could carry similar energies as the $\beta$ and $\gamma$ particles from double beta decays to excited states. These two isotopes are $\beta^{-}$ emitters from the $\mathrm{^{238}U}$ and $\mathrm{^{232}Th}$ radioactive decay chains, respectively, with $Q$-values of 3.27 and 4.99 MeV.

\begin{figure}[htpb]
  \subfloat[M{\o}ller scattering]{\includegraphics[width=0.31\textwidth]{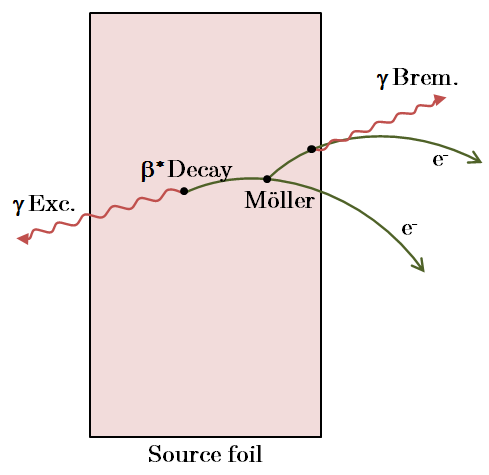} \label{fig:internal-1}}\hfill%
  \subfloat[Internal conversion] {\includegraphics[width=0.34\textwidth]{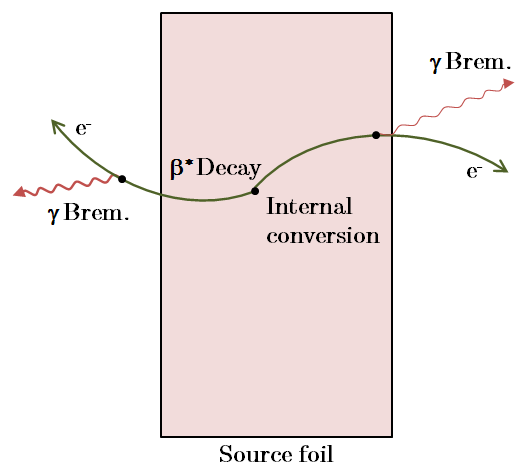} \label{fig:internal-2}}\hfill%
  \subfloat[Compton scattering] {\includegraphics[width=0.31\textwidth]{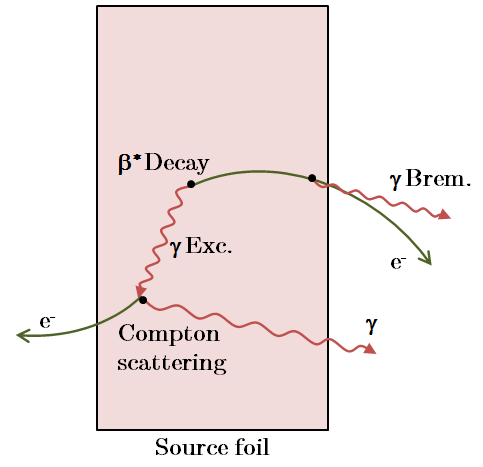} \label{fig:internal-3}}\\
  \caption{Mechanisms producing 2e2$\gamma$ events from internal contamination of $\beta$ emitters inside the source foils. $\beta$-decay to excited state followed by M{\o}ller scattering and bremsstrahlung (3(a)), $\beta$-decay to excited state with internal conversion and double bremsstrahlung  (3(b)), $\beta$-decay to excited state followed by Compton scattering and bremsstrahlung (3(c)).}
  \label{fig:internal}
\end{figure}

The main background contribution came from contamination in the
source foils introduced during isotope production and residual
contamination after isotope purification or during the foil
production. This is described as \emph{internal} contamination. In this case, those
$\beta$ emitting isotopes could produce two electrons coming from the
same vertex via $\beta$-decay with internal conversion, $\beta$-decay
followed by M{\o}ller scattering or $\beta$-decay to an excited state
with a Compton scattering of the emitted photon. From these
mechanisms, additional $\gamma$-rays could be produced by bremsstrahlung or
from a decay to an excited state as presented in Figure
\ref{fig:internal}. Prior to their installation, the activity of
$^{82}$Se foils in $^{214}$Bi and $^{208}$Tl had been measured by low
background gamma spectrometry using HPGe detectors. These small
contaminations had been also measured and cross-checked by the NEMO-3
detector itself thanks to its capability to measure own
background. In NEMO-3, the $^{214}$Bi contamination of the foils could
be studied by looking for the so called BiPo effect using $^{214}$Bi
and $^{214}$Po sequential decay events. The $\beta$-decay of
$\mathrm{^{214}Bi}$ is followed by the $\alpha$-decay of
$\mathrm{^{214}Po}$ with a half-life of 164.3 $\mu$s. The analysis
channel used to study such events was the 1e1$\alpha$(n)$\gamma$
channel. $\mathrm{^{208}Tl}$ decays exclusively to excited states and
emits mostly 2 or 3 $\gamma$-rays (99.9\%). Its contamination was thus
measured through the 1e2$\gamma$ channel with a high $\gamma$-rays
efficiency (about 50\% at 1 MeV). Results and comparison of the $^{214}$Bi
and $^{208}$Tl activities for $^{82}$Se source foils using HPGe and NEMO-3 data
are presented in Table \ref{tab:InternalHPGe} (including some other minor background isotopes \cite{Arnold:2018se}).

Finally, $2\nu\beta\beta$ decay to the ground state was also considered as a background source for $2\nu\beta\beta$ decay to excited states. When two electrons were produced, two extra $\gamma$-rays could be emitted via bremsstrahlung. A 2e2$\gamma$ event was thus detected while excited states were not involved.

\begin{figure}[htpb]
  \subfloat[Double Compton scattering]{\includegraphics[width=0.31\textwidth]{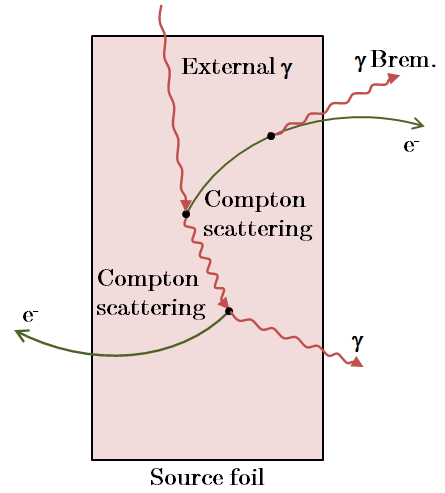} \label{fig:external-1}}\hfill%
  \subfloat[Pair production] {\includegraphics[width=0.34\textwidth]{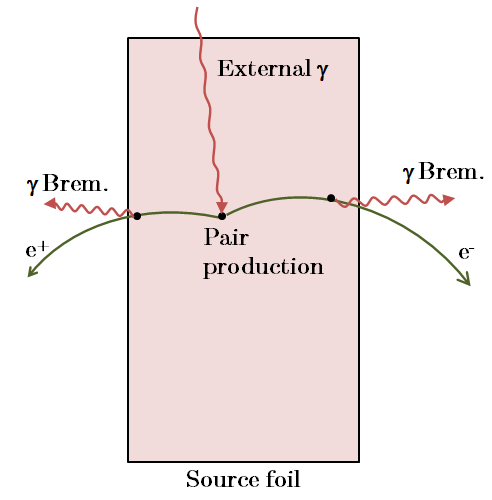} \label{fig:external-2}}\hfill%
  \subfloat[M{\o}ller diffusion] {\includegraphics[width=0.31\textwidth]{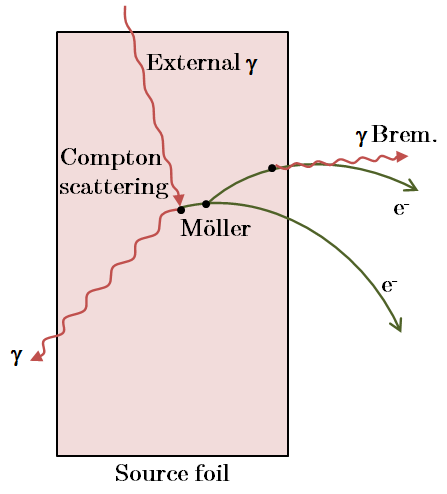} \label{fig:external-3}}\\
  \caption{Mechanisms producing 2e2$\gamma$ events from external
    contamination of the NEMO-3 detector emitting a $\gamma$-ray
    interacting inside the source foil. Double Compton scattering of
    the external $\gamma$-ray with bremsstrahlung in 4(a), pair
    production from the external $\gamma$-ray with double
    bremsstrahlung effects and poor reconstruction of the positron in
    4(b), Compton scattering of the external $\gamma$-ray followed by
    a M{\o}ller scattering of the electron and a bremsstrahlung in 4(c).}
  \label{fig:external}
\end{figure}

\begin{table}[htb]
\begin{center}
  \begin{tabular}{ c c c }
    Isotope & NEMO-3 (mBq/kg) & HPGe (mBq/kg)\\
    \hline
    $^{214}$Bi        & $1.62 \pm 0.05$ & $1.2 \pm 0.5$\\
    $^{208}$Tl        & $0.39 \pm 0.01$ & $0.40 \pm 0.13$\\
    $^{234\text{m}}$Pa & $16.7\pm 0.1$ & $<18$\\
    $^{40}$K          & $58.9\pm 0.2$ & $55 \pm 5$\\
  \end{tabular}
  \caption{Results of the contamination measured in the $^{82}$Se source
    foils by using independently NEMO-3 and HPGe data. All uncertainties
    are of statistical origin only, given at the $1\sigma$ level.  The limit shown is at the $2\sigma$ level. The activities of $^{214}$Bi and $^{208}$Tl are derived from this independent analysis and are consistent with the ones already published in        \cite{Arnold:2018se}.}
  \label{tab:InternalHPGe}
\end{center}
\end{table}
In addition to the \emph{internal} contamination of the source foils,
radioactivity from other components of the detector can produce
background events, leading to $\gamma$-rays. These $\gamma$-rays then interact with the source
foil and two electrons coming from the same vertex can then be
reconstructed if there is either pair production with
misreconstruction of the positron track, double Compton scattering or
simple Compton scattering followed by M{\o}ller scattering of the
produced electron. In the case of pair production, there can be
annihilation of the positron which produces two photons. Considering
that $\gamma$-ray interactions are involved in all those mechanisms,
they have to be taken into account in the search of the $\mathrm{2 \nu
  \beta \beta}$ and $\mathrm{0 \nu \beta \beta}$ decay to the
$\mathrm{0^+_1}$ excited state, with 2 $\gamma$-rays emitted in cascade.
The different processes responsible for background
production are described in Figure \ref{fig:external}.  The
radioactivity of these external elements was first screened by low
background $\gamma$-spectrometry. Also, when background isotopes produce a
$\gamma$-ray, it can interact close to the surface of a calorimeter
block and produce an electron. The latter crosses the whole wire
chamber including the source foil. The initial $\gamma$-ray can also
deposit energy in the calorimeter before interacting with the source
and producing an electron. The contamination of external elements can
thus be measured through two channels : crossing electron or
($\gamma,e$) external, i.e. Compton scattering in a scintillator block,
producing a $\gamma$-ray energy deposit,
followed by a Compton scattering in the source foil, emitting an
electron detected in another scintillator block. An \emph{external} background model was produced and can be found in \cite{Argyriades:2009vq}.

A specific \emph{external} background is the \emph{radon} background. It comes from $^{222}$Rn, a gaseous
isotope in the $^{238}$U chain.
$^{222}$Rn can be introduced via several mechanisms including emanation from detector
materials, diffusion from laboratory air through detector seals or contamination of the wire chamber gas.
This is only possible because of its long half-life of 3.82 days. Once
inside the detector, mainly positive ions are produced from the radon
decays. Because of their charge, they can drift and be deposited on the source foils or tracker wires. There, they decay into $^{214}$Bi near the source material. This contamination can then be observed through the 1e1$\alpha$(n)$\gamma$ channel.

For the first 18 months of data-taking, there was a relatively high level of $^{222}$Rn inside the detector. To reduce it, an anti-radon tent  was built around the detector reducing the radon level
inside the wire chamber volume by a factor $\sim 6$ \cite{NEMO32015}. The higher radon activity data-taking period is referred to as Phase 1 and the lower activity period that came after as Phase 2.

Both data and Monte Carlo simulations (MC) of signal and background
are processed by the same reconstruction algorithm. The DECAY0 event
generator \cite{DECAY0} is used for generation of initial kinematics and
particles are tracked through a detailed GEANT3 based detector
simulation \cite{GEANT3}.

\section{Gamma tracking technique}
In most double-beta-decay experiments, a crucial aspect is to precisely measure the energy of the particles. Using the unique combination of tracking and calorimetry, NEMO-3 extracts other observables (angle between two electrons, track curvature, vertex position...) allowing a good discrimination of background and signal events. In addition, one of the most important features is the measurement of the time of flight of the particles inside the detector.

When looking for double-beta decays, selecting events with two
electrons from the same vertex is not a strong enough criterion as
seen in section \ref{sec:bckg}. The time of flight measurement thus
allows to reject external events by testing two hypothesis : the event
has an \emph{internal} or an \emph{external} origin. This test is made
possible in NEMO-3 by the knowledge of the particle track length,
energy, time of flight and the energy and time resolution
($\sigma_{t}$) of the calorimeter. It can be conducted for charged
particles for which tracks are reconstructed but also for $\gamma$-rays coming from the same vertex.

Time of flight for electrons is thus an essential parameter when
looking for double-beta decay. The next section will describe its
measurement in NEMO-3 before a new method for measuring $\gamma$-ray
time of flight is presented. The latter is crucial since a more
accurate description of events containing $\gamma$-rays and a higher
sensitivity to these events will improve the efficiency and precision
in the search for decays to excited states.

\subsection{Time of flight calculation}
\label{sec:TOF}
\begin{figure}[htpb]
  \subfloat[]{\includegraphics[width=0.5\textwidth]{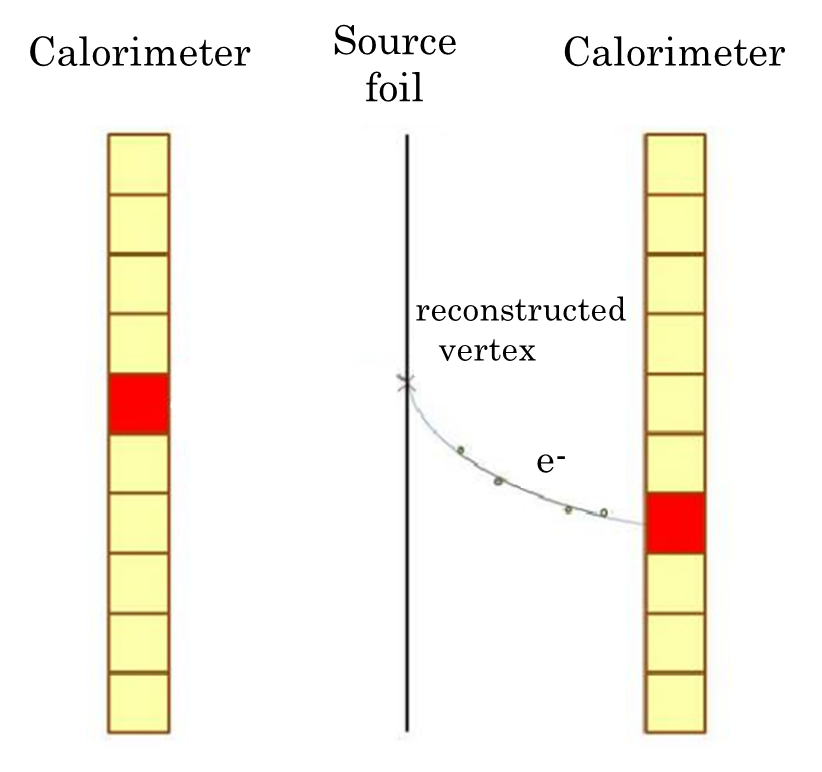} \label{fig:evt-1}}\hfill%
  \subfloat[] {\includegraphics[width=0.5\textwidth]{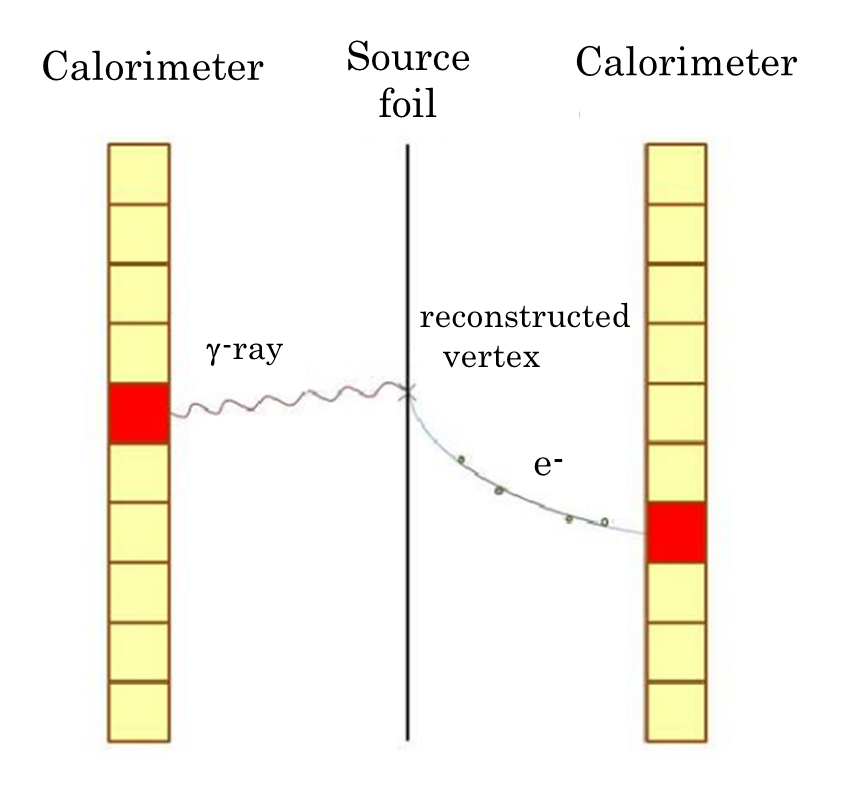} \label{fig:evt-2}}\\
  \caption{Event sketch with track reconstruction (5(a)) and scintillator association (5(b)). The reconstruction defines a $\gamma$-ray as energy deposit in a scintillator without any associated track. It can link it to the vertex in Figure 5(b).}
  \label{fig:evt-A}
\end{figure}

In order to construct an hypothesis on the time ordering of an event,
some energy must be deposited in at least two calorimeter blocks and
one particle track or more must be reconstructed inside the wire
chamber. This track also has to be associated to one of the
calorimeter hits. The other calorimeter hit with no associated track
is identified as a $\gamma$-ray. Figure \ref{fig:evt-A} illustrates an
event sketch in NEMO-3 with an electron (one reconstructed track with one calorimeter hit) and a $\gamma$-ray (only a calorimeter hit) coming from the same vertex.

Before making any time of flight calculation for an event, two hypotheses must be considered : \emph{internal} or \emph{external} origin. Theoretical times of flight between the vertex and the calorimeter hit ($t^{th}$) that should be measured by the calorimeter (for each block hit) are then calculated for both hypotheses. The sum (\emph{external} origin) or difference (\emph{internal} origin) of these theoretical times is compared to the difference between times actually measured by the calorimeter ($t^{exp}$). If the given hypothesis is favoured, then the difference noted $\triangle t_{hyp}$ must be close to zero, taking into account the time resolution of NEMO-3.

Considering the example presented in Figure 5(b), these differences for both hypotheses are expressed by the following equations :

\begin{equation}
\triangle t_{int}=(t_{e}^{th}-t_{\gamma}^{th})-(t_{e}^{exp}-t_{\gamma}^{exp})
\label{eq:int}
\end{equation}
\begin{equation}
\triangle t_{ext}=(t_{e}^{th}+t_{\gamma}^{th})-(t_{e}^{exp}-t_{\gamma}^{exp})
\label{eq:ext}
\end{equation}

Nevertheless, the calculation of $\triangle t_{hyp}$ is only a preliminary analysis. A more advanced study is based on the probability of time of flight and needs to take into account the uncertainties on theoretical and measured times. Thus the $\chi^{2}$ method is used as described by :

\begin{equation}
\chi_{hyp}^{2}=\frac{\triangle^{2}t_{hyp}}{\sigma_{tot}^{2}},
    \end{equation}

\noindent where $\sigma_{tot}^{2}$ is the quadratic sum of all uncertainties affecting time measurements or calculations. These are the uncertainties on track lengths (for charged particles), path lenghts (for $\gamma$-ray), measured energies (due to calorimeter energy resolution) and times (due to calorimeter time resolution).

When considering \emph{external} or \emph{internal} events, as in
section \ref{sec:double-b}, the selections will be based on the
chi-squared probabilities for the respective hypotheses.

\subsection{Gamma tracking}
\label{sec:GT}

Another type of time of flight calculation is possible considering only the trajectory of photons. Because of the thickness of NEMO-3 scintillators, $\gamma$-rays do not always deposit all their energy inside a single block. One photon can deposit part of its energy in a calorimeter block after Compton scattering, then hit another one and potentially more. Gamma tracking  is an original and powerful analysis tool developed recently \cite{theseHugon} in order to take this effect into account and reconstruct the complete trajectory of $\gamma$-rays inside the detector, with each step from one scintillator to the next.

When a single $\gamma$-ray is produced inside a source foil with one
or more charged particles and hits several scintillators, a few PMTs
measure energies without associating them to reconstructed
tracks. Figure 6(a) describes the approach presented in the previous
section, where every unassociated hit is considered as having a
different origin. Here, the second unassociated block is neither
\emph{internal} nor \emph{external} and the event is excluded when
selecting events for the 2e1$\gamma$ channel. Using gamma tracking,
the same event can be properly reconstructed as shown in Figure 6(b) :
the second unassociated hit is paired with the first one under the
assumption of Compton scattering and the event satisfies the 2e1$\gamma$ channel conditions.

\begin{figure}[htpb]
  \subfloat[]{\includegraphics[width=0.5\textwidth]{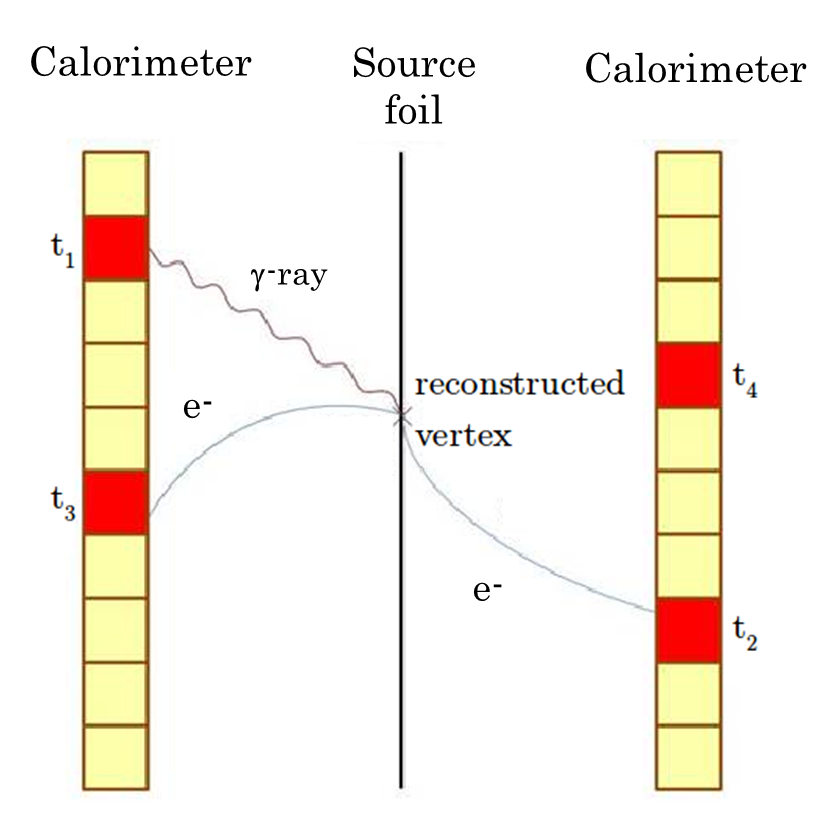} \label{fig:evt-3}}\hfill%
  \subfloat[] {\includegraphics[width=0.5\textwidth]{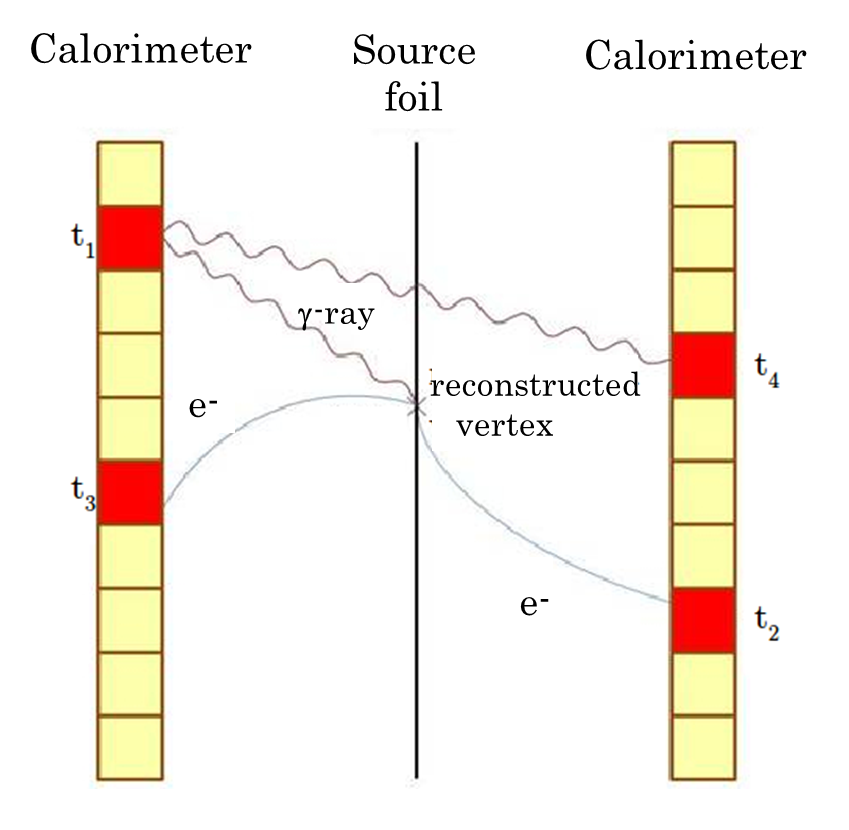} \label{fig:evt-4}}\\
  \caption{Example of an event reconstruction without using gamma
    tracking (6(a)). Only one of the two scintillators not associated
    to a track is consistent with the \emph{internal} hypothesis, the
    other is neither  \emph{internal} nor  \emph{external}. The same
    event is reconstructed with gamma tracking (6(b)), the second
    scintillator can be associated to the first one under the
    assumption of Compton scattering.}
  \label{fig:evt-B}
\end{figure}

In this example, the complete reconstruction of the photon can be done
with only one time of flight probability calculation : between the two
scintillators not associated to any track. However, when events
include several unassociated calorimeter blocks, every combination has
to be taken into account and evaluated. In that case, before making a
complete calculation, the probability of time of flight is determined
for each pair of blocks in the event, with once again the $\chi^{2}$
method.  All the pairs are then combined to extract all possible
topologies, each associated to a combined time of flight probability, using the equation :

\begin{equation}
\chi_{GT_{tot}}^{2}=\sum_{n=1}^{n=m}\chi_{GT}^{2}(Block_{n-1}Block_{n}),
\end{equation}

\noindent with $m$ the total number of blocks involved in the chain of hit scintillator by a single $\gamma$-ray, $Block_{n-1}Block_{n}$ a pair of calorimeter blocks and $\chi_{GT}^{2}(Block_{n-1}Block_{n})$ the $\chi^{2}$ value calculated for each pair. The main drawback of this method is the computation time. To limit this effect, two additional conditions are applied : requiring an energy threshold of 150 keV for the energy deposit in each calorimeter block and only taking into account the probabilities greater than $0.1 \%$ for any combination.

Once all calculations have been performed, the topology with the
highest probability is considered the most likely. This combination
defines the number of photons in the event and their trajectories. The
gamma tracking technique is thus key to the study of double-beta decays to excited states.

\subsection{Validation of gamma tracking using calibration sources}
\label{sec:source}

During the data taking phase of NEMO-3, several calibration runs using three point-like $^{232}$U radioactive sources, labelled 1, 2 and 3, were conducted. Their activities were measured through $\gamma$-spectrometry (HPGe detectors) and are given in Table \ref{tab:act-tl}, column 2. These sources are especially well suited for gamma tracking studies since they decay to the $^{228}$Th nucleus which belongs to the natural $^{232}$Th radioactive decay chain. At the end of the chain, it produces a $^{208}$Tl nucleus which is a $\beta^{-}$ emitter producing at least two $\gamma$-rays : e.g. 2.615 and 0.583 MeV.

\begin{figure}[htpb]
%  \subfloat[]{\includegraphics[width=0.5\textwidth]{GT_size.png} \label{fig:GT_size}}\hfill%
   {\includegraphics[width=1.\textwidth]{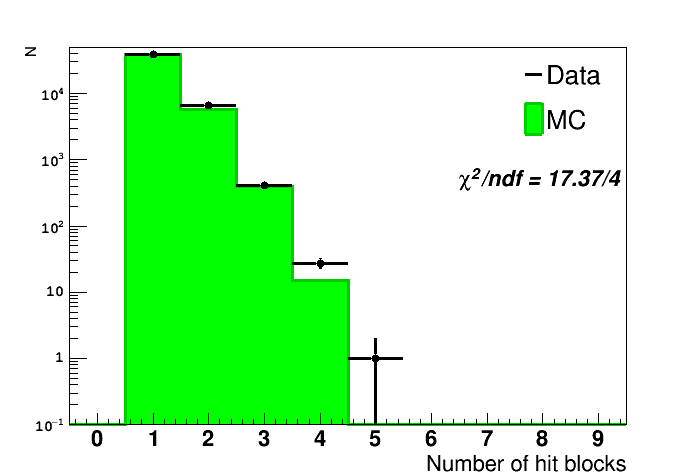} \label{fig:GT_size_log}}\\
  \caption{Number of scintillator blocks hit by a single $\gamma$-ray according to the path reconstruction calculated with the gamma tracking method, logarithmic scale. Data were acquired using $\gamma$-rays from the $^{232}$U radioactive sources and are compared to Monte-Carlo simulations.}
  \label{fig:gt-rebound}
\end{figure}

We measured the activities of the $^{232}$U sources using NEMO-3
analysis with and without gamma tracking. We can then compare the
results with the activities measured by HPGe detectors. The main
objectives are to confirm that the use of the gamma tracking method
improves the signal efficiency and reduces the systematics. Several
criteria are defined to only select events involving one electron and
two $\gamma$-rays ($\mathrm{1e2\gamma}$) since $99.8 \%$ of $^{208}$Tl
decays produce these three particles. Using Monte-Carlo simulations,
the efficiency with gamma tracking is determined to be $1.16 \%$ for
this topology (compared to $0.92 \%$ without gamma tracking) while
53082 data events are selected for source 3 with an acquisition time
of 107.6 hours. The same analysis without gamma tracking, conducted on
the same data sample, selected only 38956 events. About $27 \%$ more
events were thus selected using gamma tracking, proving that part of
the events involving Compton scattering are recovered, thus improving the efficiency. Figure \ref{fig:gt-rebound} illustrates the number of scintillator blocks hit by a single $\gamma$-ray according to the path reconstruction calculated with the gamma tracking method. A reasonable agreement is obtained between data from the $^{232}$U radioactive sources and Monte-Carlo simulations. 

Futhermore, the $^{232}$U sources activities obtained with gamma tracking are presented in Table \ref{tab:act-tl} where they are compared to activities obtained without the use of gamma tracking and to the $\gamma$-spectrometry measurements.
The interaction of $\gamma$-rays may induce low energy deposits in the bulk of the scintillator block. The energy response of the calorimeter does not take into account the interaction point in the scintillator block so the effect of the energy threshold may be difficult to simulate. However, the main observation is that activities measured using gamma tracking are more consistent with the $\gamma$-spectrometry results.

\begin{table}
\begin{centering}
\medskip{}

\par\end{centering}

\begin{centering}
\begin{tabular}{|c|c|c|c|c|c|}
\hline 
$^{232}$U & HPGe (Bq) & No GT act. (Bq) & $\Delta_{noGT}$ ($\%$) & GT act. (Bq) & $\Delta_{GT}$ ($\%$)\tabularnewline
\hline 
\hline 
1 & $7.79\pm0.04\pm0.21$ & $6.56\pm0.08$ & 15.8 & $6.98\pm0.07$ & 10.4\tabularnewline
\hline 
2 & $15.91\pm0.09\pm0.43$ & $13.92\pm0.13$ & 12.5 & $14.88\pm0.11$ & 6.5\tabularnewline
\hline 
3 & $32.76\pm0.17\pm0.89$ & $30.00\pm0.17$ & 8.4 & $32.11\pm0.14$ & 2.0\tabularnewline
\hline 
\end{tabular}
\par\end{centering}

\caption{Comparison of the $^{232}$U sources activities measured by respectively $\gamma$-spectrometry
(HPGe detector) and NEMO-3 analysis without and with the gamma
tracking technique using the $1e2\gamma$ topology. Uncertainties in column 2 are respectively statistics and systematics. Columns 4 and 6 present the relative differences between HPGe and analysis activities (without and with gamma tracking).}
\label{tab:act-tl}

\medskip{}
\end{table}

However, activities measured through the analysis with gamma tracking are consistently lower than $\gamma$-spectrometry values. This difference is used as a way to estimate the systematic uncertainty induced by the use of the gamma tracking technique. The difference for sources 1, 2 and 3 are respectively $10.4 \%$, $6.5 \%$ and $2.0 \%$ as reported in Table \ref{tab:act-tl}. As a conservative approach, the systematic uncertainty is considered to be $10 \%$.

\section{Double beta decay to the excited states}
\label{sec:double-b}

\subsection{Two neutrino double-beta decay to $0_{1}^{+}$ excited state}
\label{sec:2n2b}

\begin{figure}[htpb]
  \subfloat[Phase 1]{\includegraphics[width=0.5\textwidth]{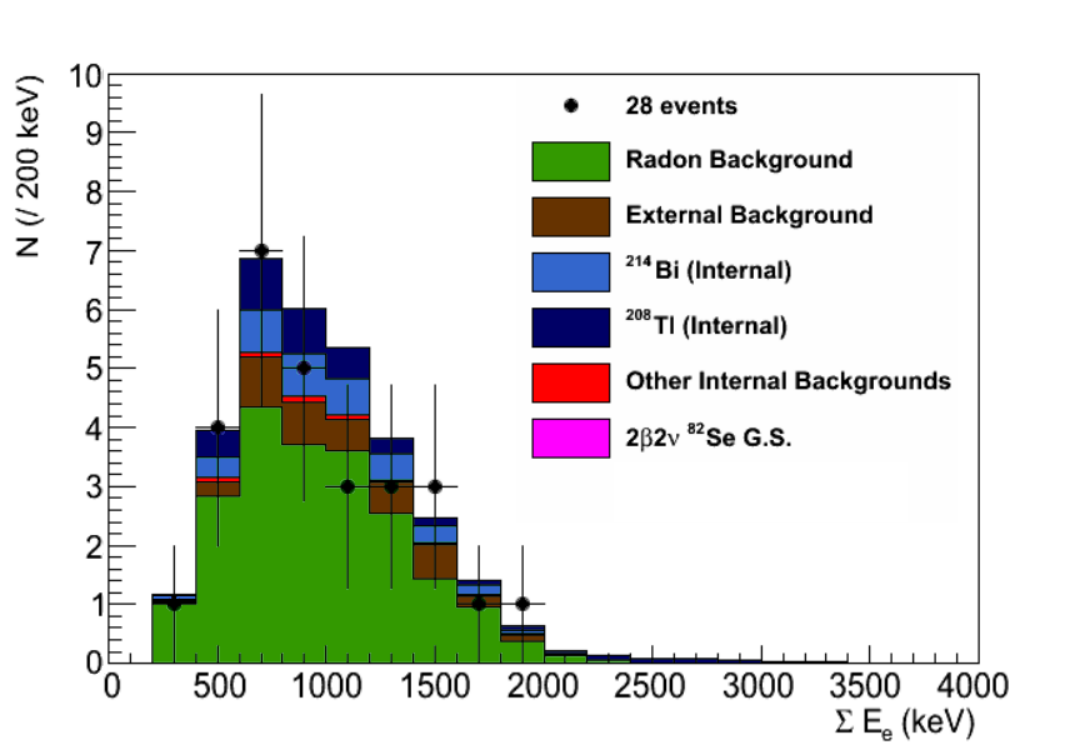} \label{fig:Ee_P_Ph1_2n}}\hfill%
  \subfloat[Phase 2] {\includegraphics[width=0.5\textwidth]{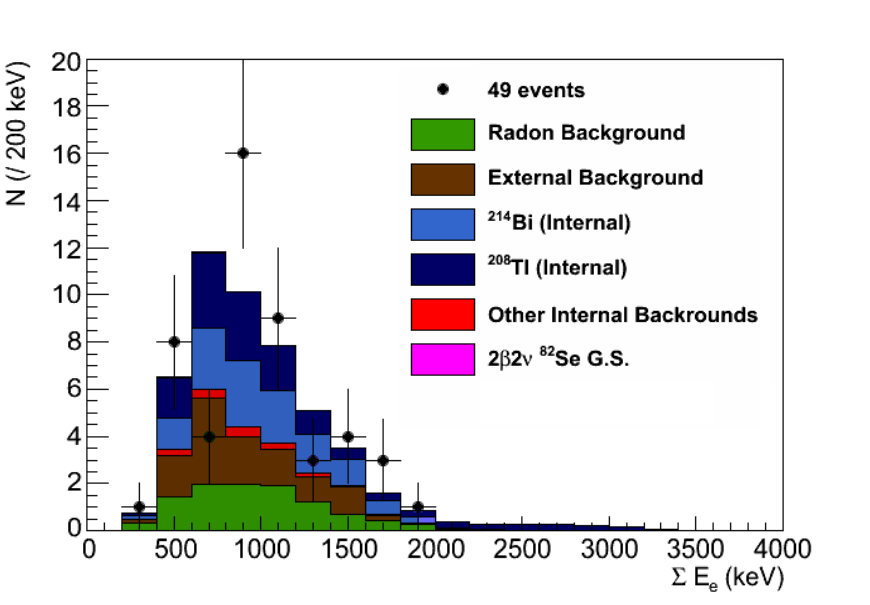} \label{fig:Ee_P_Ph2_2n}}\\
  \caption{Sum of the electron energies distributions in the
    2e2$\gamma$ channel after the preselection criteria described in
    the text are applied, for Phase 1 in Figure 8(a) and Phase 2 in Figure 8(b). Data are compared to the MC prediction for the different backgrounds. The background coming from the $2\nu\beta\beta$ $\mathrm{^{82}Se}$ decay to g.s. is completely negligible and thus not visible in the two plots.} 
  \label{fig:Ee_P_2n}
\end{figure}

As mentioned in Section \ref{sec:intro}, $\beta\beta$ decays to the
$0_{1}^{+}$ excited state consist in the simultaneous emission
(compared to the NEMO-3 time resolution) of two $\beta$ and two $\gamma$ particles. In order to select 2e2$\gamma$ events, several criteria are applied to distinguish them from background events. The candidate events must contain two electron tracks, originating from the $^{82}$Se source foil, each with an energy deposit
 greater than $150~\mbox{keV}$. The distance between the tracks' intersections with
the foil should fulfill $\Delta_{XY}$ less than 4 cm (perpendicular to the wires) and 
$\Delta_{Z}$ less than 8 cm (parallel to the wires)	 so they can be
considered to have a common vertex. Two $\gamma$-rays must be
reconstructed using the gamma tracking technique,  each with a total
energy greater than
$150~\mbox{keV}$. The timing of the calorimeter hits for electrons and $\gamma$-rays must be consistent with an \emph{internal} event
defined as those particles simultaneously emitted from their common vertex in the $\mathrm{^{82}Se}$ foil. There should be no $\alpha$-particle
tracks and no extra reconstructed $\gamma$-rays in the event.  77 data events were selected from a total of 897,409,450 in the selenium sectors for the selected runs. Figure \ref{fig:Ee_P_2n} shows that this number is compatible with the number of background events expected when using these criteria, as well as the energy distribution of both electrons for data events and background. Using these criteria, the efficiency for the expected signal is $0.078 \%$.

These preselection criteria can be applied when looking for events including two internal electrons and $\gamma$-rays. In order to be more specific to the $2\nu\beta\beta$($0_{gs}^{+}\rightarrow0_{1}^{+}$) decay, an optimisation is made considering the energies of the four particles for this decay.
The first energies to be optimized are the individual energies of both electrons labelled $\mathrm{E_{e~min}}$ and $\mathrm{E_{e~max}}$. Both energies for each event are displayed using two-dimensional histograms for signal and total background, obtained from MC simulations as shown in Figures 9(a) and 9(b). For each bin of this two-dimensional histogram, the local statistical significance (noted $N_{\sigma}^{l}$
 ) is calculated and displayed in Figure 9(c). This value is defined by the following equation :

\begin{equation}
 N_{\sigma}^{l}=\frac{S^{l}}{\sqrt{S^{l}+B^{l}}},
\end{equation}

\noindent where $S^{l}$ is the signal and $B^{l}$  the background in each bin.  
The signal is given by the $2\nu\beta\beta$ to $0_{1}^{+}$ state simulation with a half-life of $3\times10^{20}$
  years which is three times higher than the $2\nu\beta\beta$ decay to the ground state half-life. 

The result of the optimization procedure was tested for several Monte Carlo samples,
including $2 \nu \beta \beta$ to the $0^+_1$ state with various
half-lives. If the half-life in the sample is different from
$\mathrm{3 \times 10^{20}}$ y, the selection would not be optimal,
thus the sensitivity to the $\mathrm{2 \nu \beta \beta}$ to the
$\mathrm{0^+_1}$ excited state would be decreased. For samples with
half-lives larger than $\mathrm{3 \times 10^{20}}$ y, the optimization
procedure gives a too loose selection w.r.t optimum, increasing the background
contribution. For samples with
half-lives smaller than $\mathrm{3 \times 10^{20}}$ y, the optimization
procedure results in a too strict selection, reducing the signal efficiency.
Even if not optimal, the selection would not bias the half-life of
the sample.

%The dependency of the optimization of the selection on the half-life
%of the $2 \nu 2 \beta$ to the $0^+_1$ excited state was checked by varying the set
%half-life. This affects the sensitivity for the search of $2 \nu 2
%\beta$ to the $0^+_1$ state only in the case of too large or too small
%values of the half-life. This doesn't produce a distorsion of the
%measured half-life w.r.t. the set value.

%It was verified that the half-life calculation would
%not be affected by the half-life attributed to the simulated signal
%$S$ during the cut optimisation. To do so, the whole process was
%repeated multiple times with other half-life values.
%We observed that changing the activity of the signal only has an effect
%on the final result if it is too high or too low by affecting the
%terms in equation \ref{eq:T1/2-simple} :
%\begin{itemize}
%\item when too high, only the signal is considered in the optimisation process,
%not the background. Thus the efficiency is maximised but the energy
%cuts are looser, more background is observed and the number of excluded
%events $N_{ex}$ increases,
%\item when too low, the backgrounds are given more weight but the cuts become
%too strong, thus the efficiency decreases.
%\end{itemize}

A selection criterion is defined on $N_{\sigma}^{l}$ for the maximised total statistical significance $N_{\sigma}$ as presented in Figure 9(d). $N_{\sigma}$ is calculated over the total number of simulated signal events and expected background.

\begin{figure}[htpb]
  \subfloat[Signal]{\includegraphics[width=0.5\textwidth]{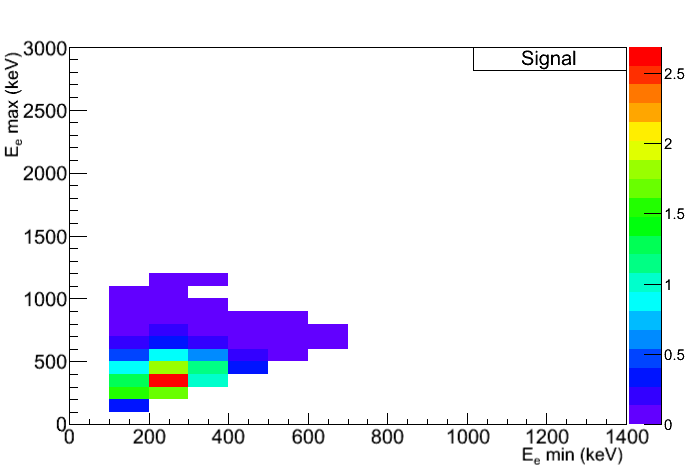} \label{fig:Ee_vs_Ee_sig}}\hfill%
  \subfloat[Background] {\includegraphics[width=0.5\textwidth]{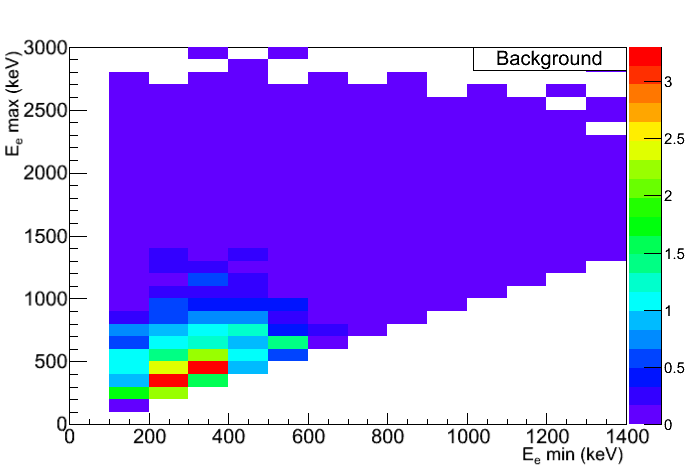} \label{fig:Ee_vs_Ee_bdf}}\hfill%
 \subfloat[Local statistical significance]{\includegraphics[width=0.5\textwidth]{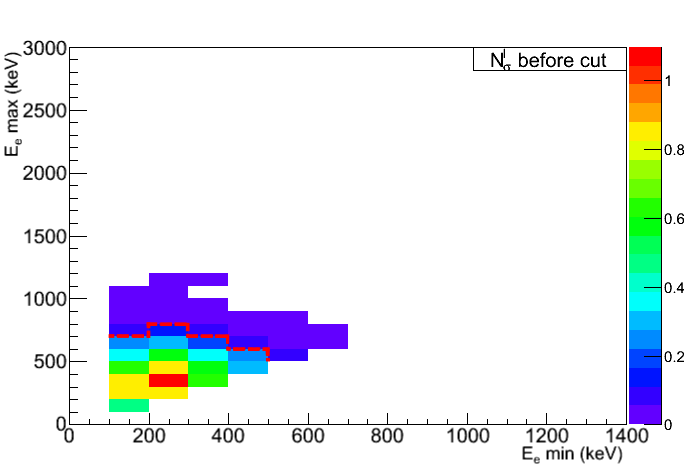} \label{fig:Ee_vs_Ee_rat}}\hfill%
  \subfloat[Total statistical significance] {\includegraphics[width=0.5\textwidth]{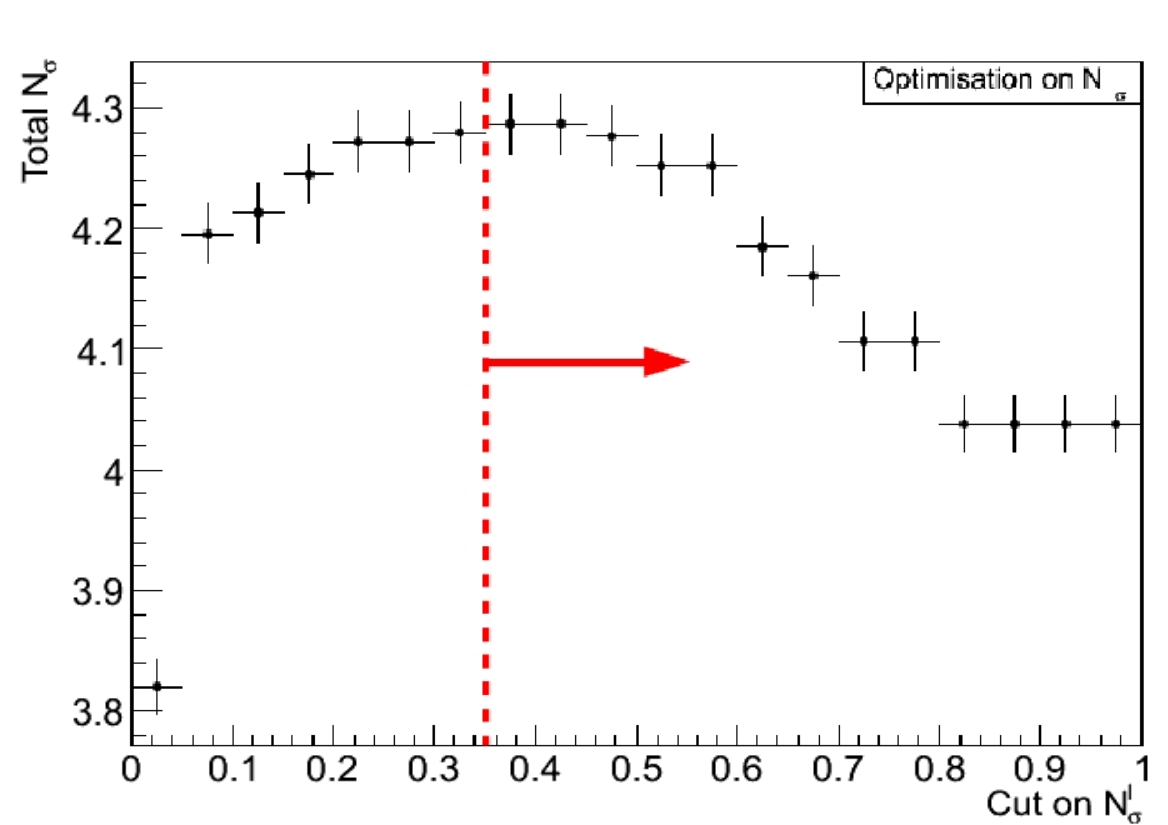} \label{fig:Ee_vs_Ee_opt}}\\
  \caption{The distributions of signal
    $2\nu\beta\beta$($0_{gs}^{+}\rightarrow0_{1}^{+}$) and background
    events from MC simulation are represented in Figures
    9(a) and 9(b) respectively, as
    a function of both electrons' individual energy. The local
    statistical significance $N_{\sigma}^{l}$ distribution for the
    $2\nu\beta\beta$($0_{gs}^{+}\rightarrow0_{1}^{+}$) transition as a
    function of both electrons' individual energy is calculated for
    each bin of the 2-D histogram and represented in Figure 9(c). The
    total statistical significance $N_{\sigma}$ as a function of a cut
    on the local significance in Figure 9(d)  allows
    the optimization of this cut (dotted red line). Selected bins with high local statistical significance in Figure 9(c) are separated from the removed ones by the dotted red line.}
  \label{fig:opti_EevsEe1}
\end{figure}

In Figure 9(a), simulations show that the signal is
stronger when the energies of the two electrons $\mathrm{E_{e~max}}$
and $\mathrm{E_{e~min}}$ are in the range of [300-400] and [200-300] keV,
respectively. This is due to their primary kinetic energies (with a
total energy shared equal or lower than $1510.2$ keV) slightly
affected by the loss of energy in the source foil and in the tracking
chamber. Concerning the background, the simulations in Figure
9(b) show that the energies of the two detected
electrons can be much higher, up to 2.7 MeV, than those for the
signal. This is due to the presence of $^{208}$Tl isotope ($Q$-value
of $4.99$ MeV) which is one of the main backgrounds. Nevertheless, the
optimization is able to remove all the events with high energy
electrons, typically greater than 1.1-1.2 MeV as illustrated in Figure 9(c). 

\begin{figure}[htpb]
\subfloat[Signal]{\includegraphics[width=0.5\textwidth]{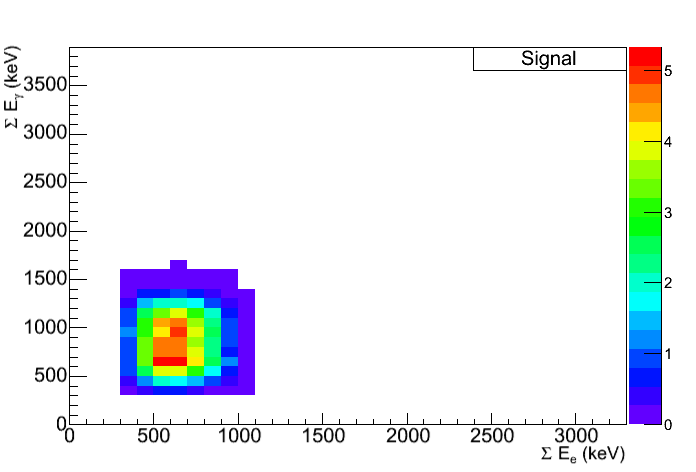} \label{fig:Ee_vs_Eg_sig}}\hfill%
  \subfloat[Background] {\includegraphics[width=0.5\textwidth]{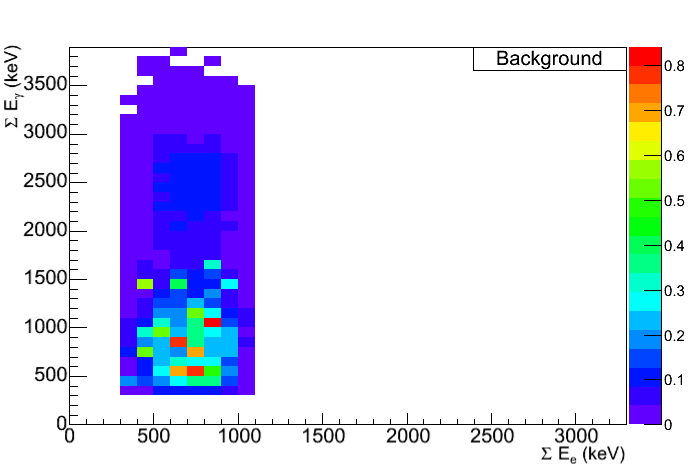} \label{fig:Ee_vs_Eg_bdf}}\hfill%
  \subfloat[Local statistical significance]{\includegraphics[width=0.5\textwidth]{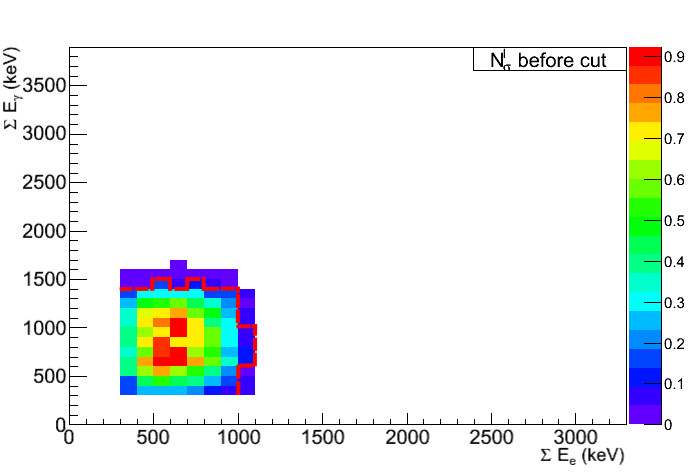} \label{fig:Ee_vs_Eg_rat}}\hfill%
  \subfloat[Total statistical significance] {\includegraphics[width=0.5\textwidth]{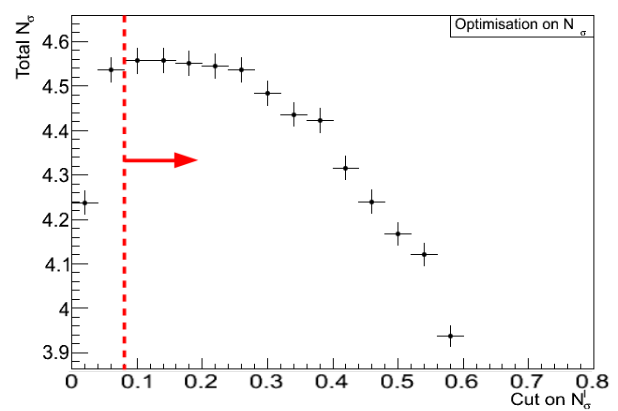} \label{fig:Ee_vs_Eg_opt}}\\
  \caption{Total electron energy vs total $\gamma$-rays energy distributions for $2\nu\beta\beta$($0_{gs}^{+}\rightarrow0_{1}^{+}$)
  signal simulation (10(a)) and background (10(b)). Local statistical significance distributions for each bin of this histogram (10(c)) with optimisation cut (dotted red line) on total statistical significance $N_{\sigma}$ presented in Figure 10(d).}
  \label{fig:opti_2n_02}
\end{figure}

\begin{figure}[htpb]
\subfloat[Signal]{\includegraphics[width=0.5\textwidth]{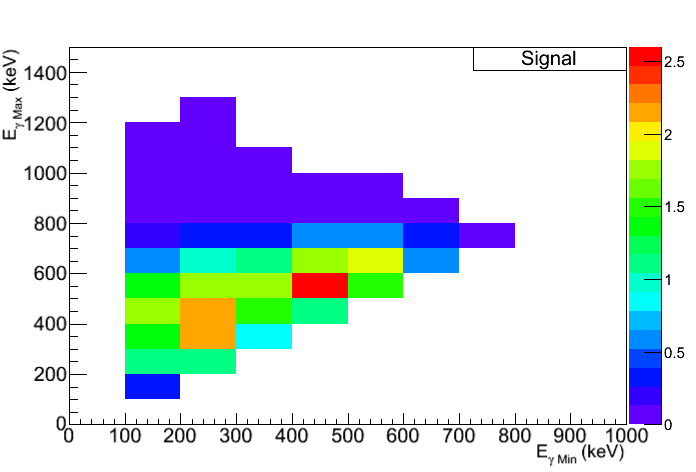} \label{fig:Eg_vs_Eg_sig}}\hfill%
  \subfloat[Background] {\includegraphics[width=0.5\textwidth]{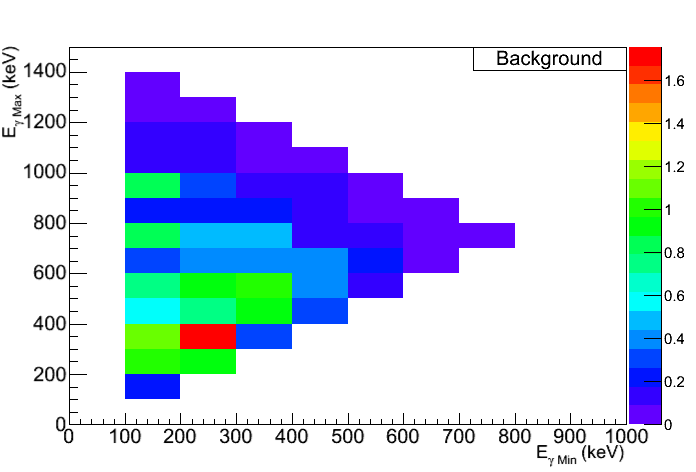} \label{fig:Eg_vs_Eg_bdf}}\hfill%
  \subfloat[Local statistical significance]{\includegraphics[width=0.5\textwidth]{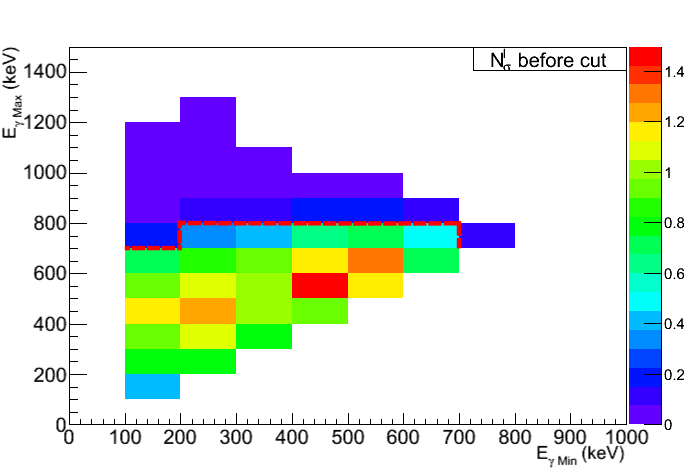} \label{fig:Eg_vs_Eg_rat}}\hfill%
  \subfloat[Total statistical significance] {\includegraphics[width=0.5\textwidth]{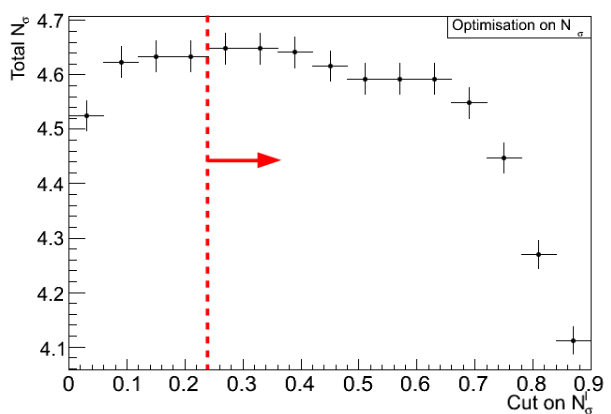} \label{fig:Eg_vs_Eg_opt}}\\
  \caption{$\gamma$-ray 1 energy vs $\gamma$-ray 2 energy distributions for $2\nu\beta\beta$
 ($0_{gs}^{+}\rightarrow0_{1}^{+}$
 ) signal simulation (11(a)) and background (11(b)). Local statistical significance distributions for each bin of this histogram (11(c)) with optimisation cut (dotted red line) on total statistical significance $N_{\sigma}$ presented in Figure 11(d).}
  \label{fig:opti_2n_03}
\end{figure}

Other selections are then made on the total electron energy and total
$\gamma$-rays energy and finally on the two $\gamma$-rays' individual energies as seen respectively in Figures \ref{fig:opti_2n_02} and \ref{fig:opti_2n_03}.
Figure \ref{fig:opti_2n_02} shows some of the features of Figure 9, whereby the total energy of $\gamma$-rays for background can be greater than $1500$ keV due to higher energy $\gamma$-rays emitted in $^{208}$Tl decays (usually $2.61$ and $0.58$ MeV). The signal simulation fits the
$2\nu\beta\beta$($0_{gs}^{+}\rightarrow0_{1}^{+}$) transition with the two electrons sharing $1512.2$ keV and two $\gamma$-rays with a total energy of $1487.7$ keV. The optimisation process then only selects events with $\gamma$-rays sharing less than $1600$ keV, taking into account the energy resolution of the detector.
Figure \ref{fig:opti_2n_03} represents the third step of optimization
and concerns individual $\gamma$-rays energies. By this stage, most of the $^{208}$Tl induced events have been removed. Simulations indicate that most of the remaining background events contain two $\gamma$-rays of [300-400] and [200-300] keV. These can be related to $^{214}$Bi, since its decay can produce a $609.3$ keV $\gamma$-ray and a lower energy one through bremsstrahlung, shown in Figure 3(a). Finally, most signal events are expected to have two $\gamma$-rays of [400-500] and [500-600] keV, corresponding to the $2\nu\beta\beta$($0_{gs}^{+}\rightarrow0_{1}^{+}$) $\gamma$-rays of $711.2$ and $776.5$ keV.

 After the complete optimization process described here, the selection efficiency for the $2\nu\beta\beta$($0_{gs}^{+}\rightarrow0_{1}^{+}$) signal calculated from MC is $0.069 \%$ with a total of 19 selected data events.

\begin{figure}[htpb]
  \subfloat[Phase 1]{\includegraphics[width=0.5\textwidth]{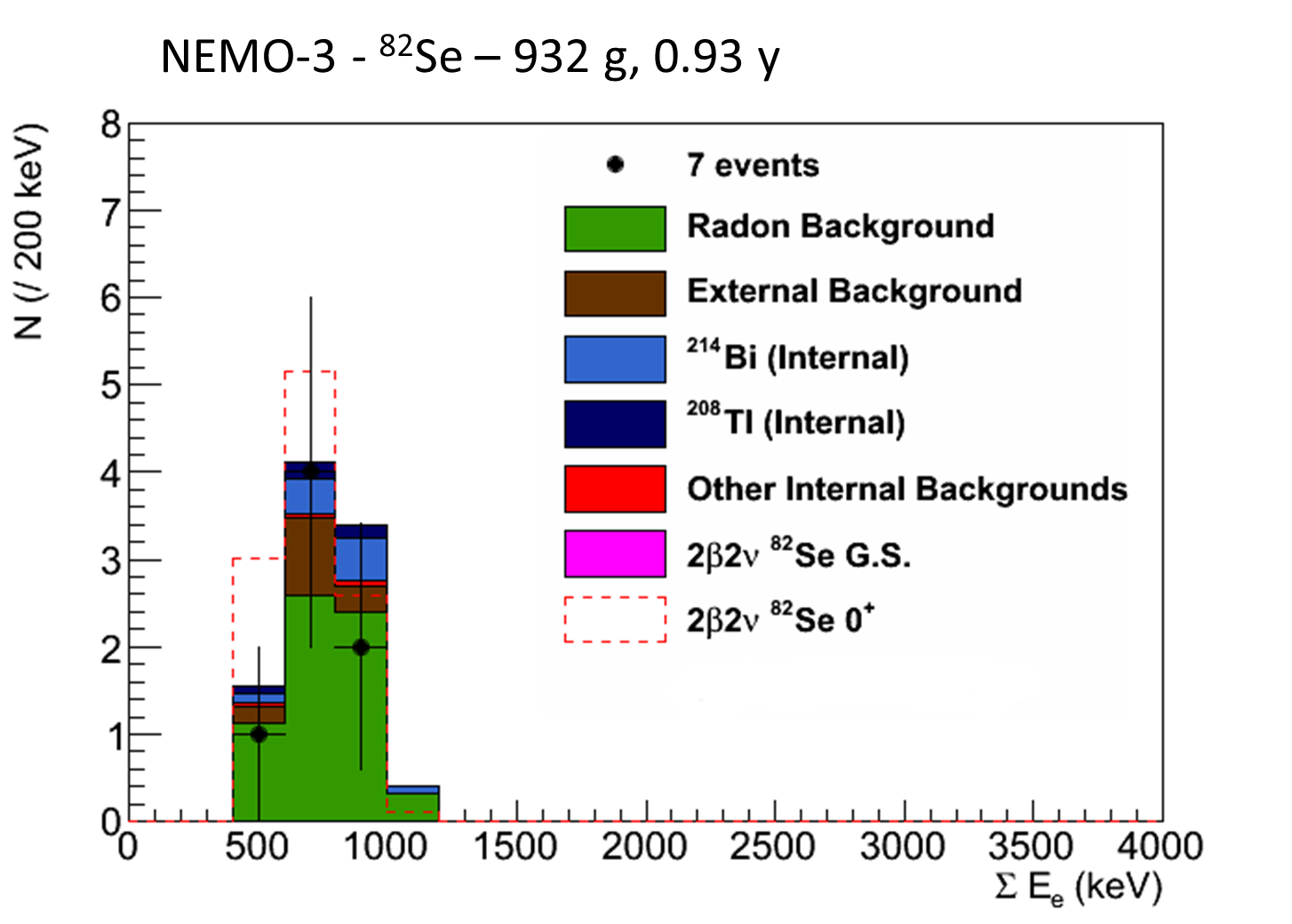} \label{fig:Ee_fin_Ph1_2n}}\hfill%
  \subfloat[Phase 2] {\includegraphics[width=0.5\textwidth]{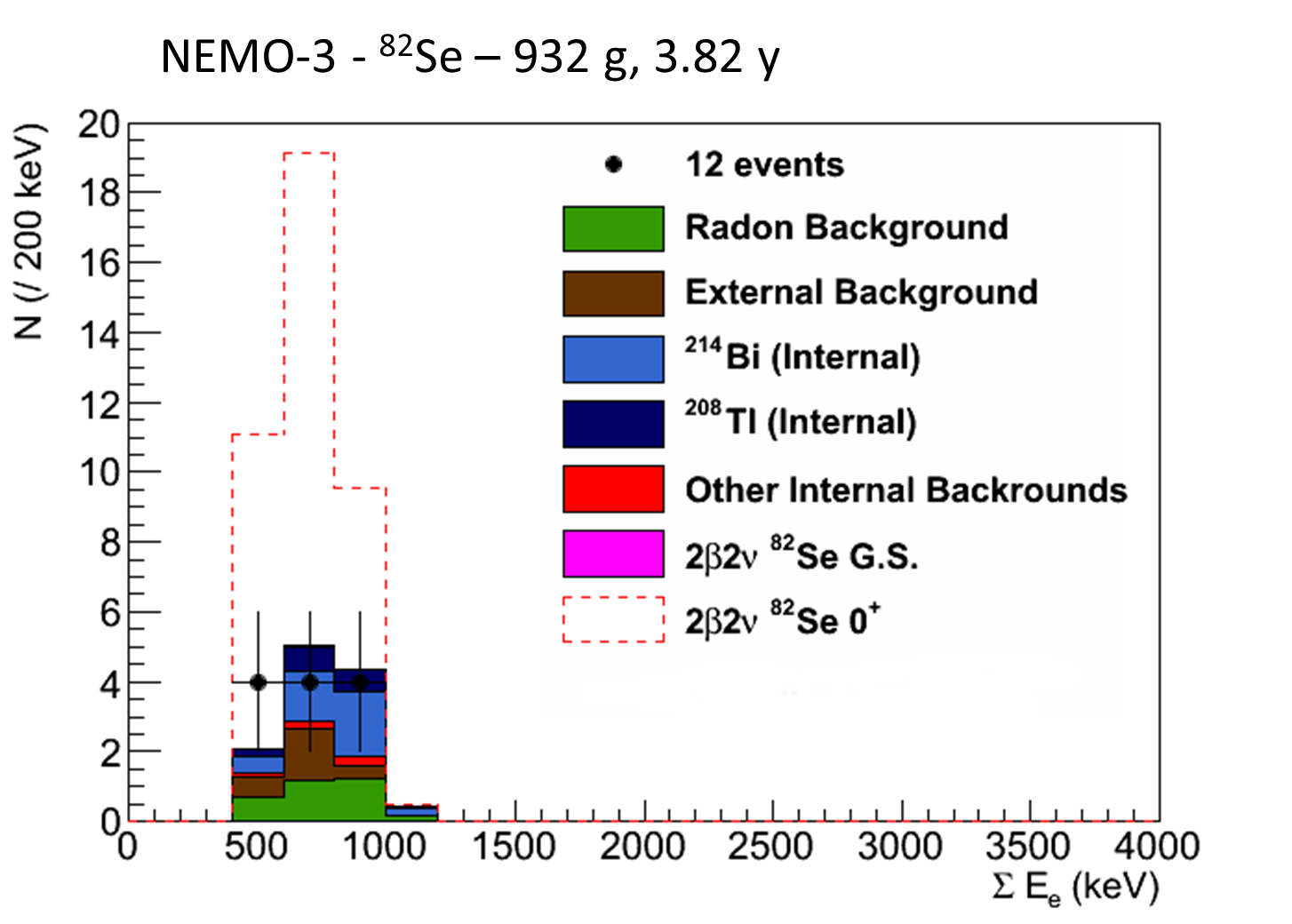} \label{fig:Ee_fin_Ph2_2n}}\\
  \caption{Total electron energy distributions after selection for the $2\nu\beta\beta$($0_{gs}^{+}\rightarrow0_{1}^{+}$) transition,
  for Phase 1 in Figure 12(a) and Phase 2 in Figure 12(b). Experimental data events are compared to the MC simulation for the different backgrounds. The dotted red line represents the simulated signal for a half-life of $3\times10^{20}$ years.}
  \label{fig:Ee_2n}
\end{figure}

\begin{figure}[htpb]
  \subfloat[Phase 1]{\includegraphics[width=0.5\textwidth]{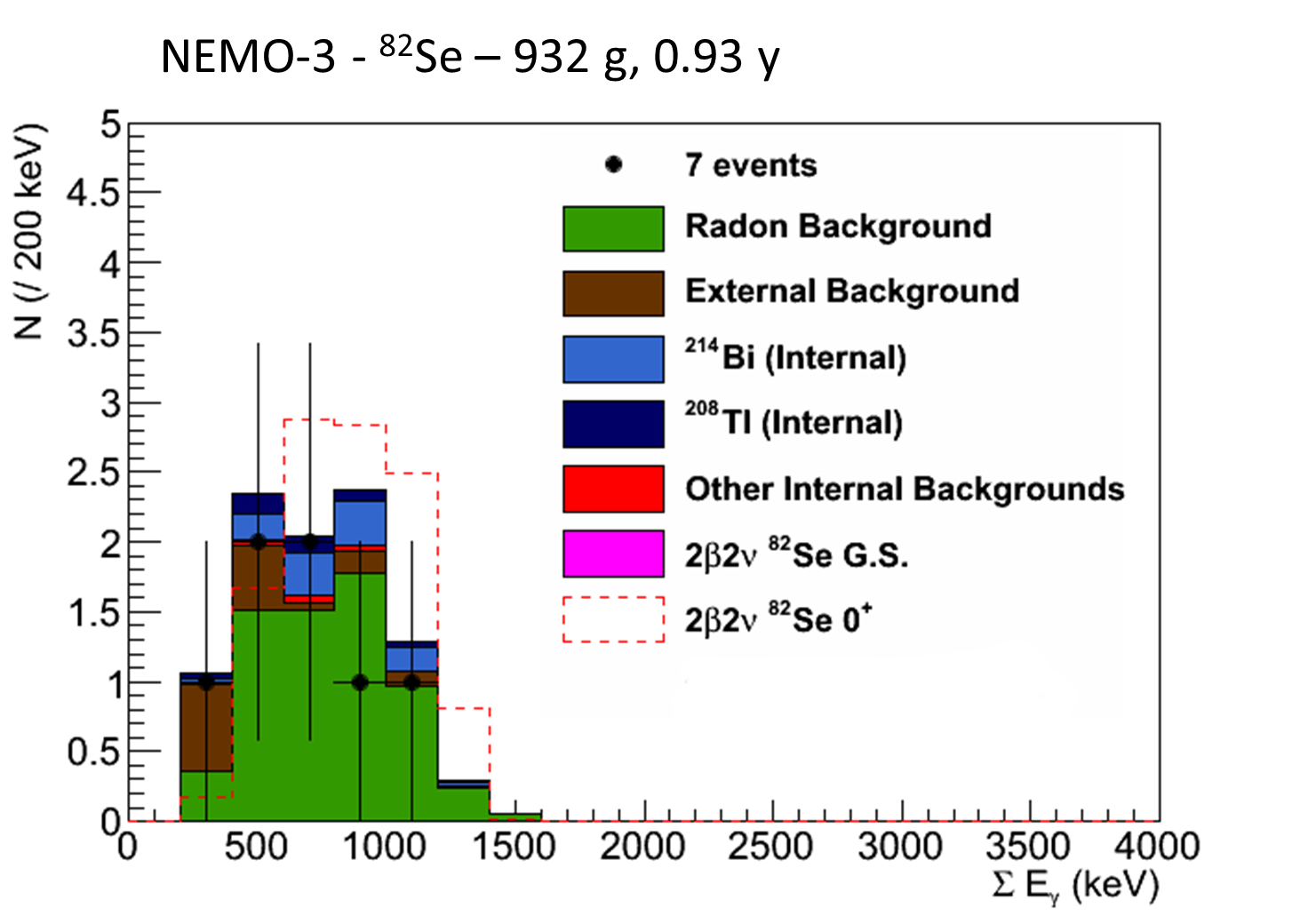} \label{fig:Eg_fin_Ph1_2n}}\hfill%
  \subfloat[Phase 2] {\includegraphics[width=0.5\textwidth]{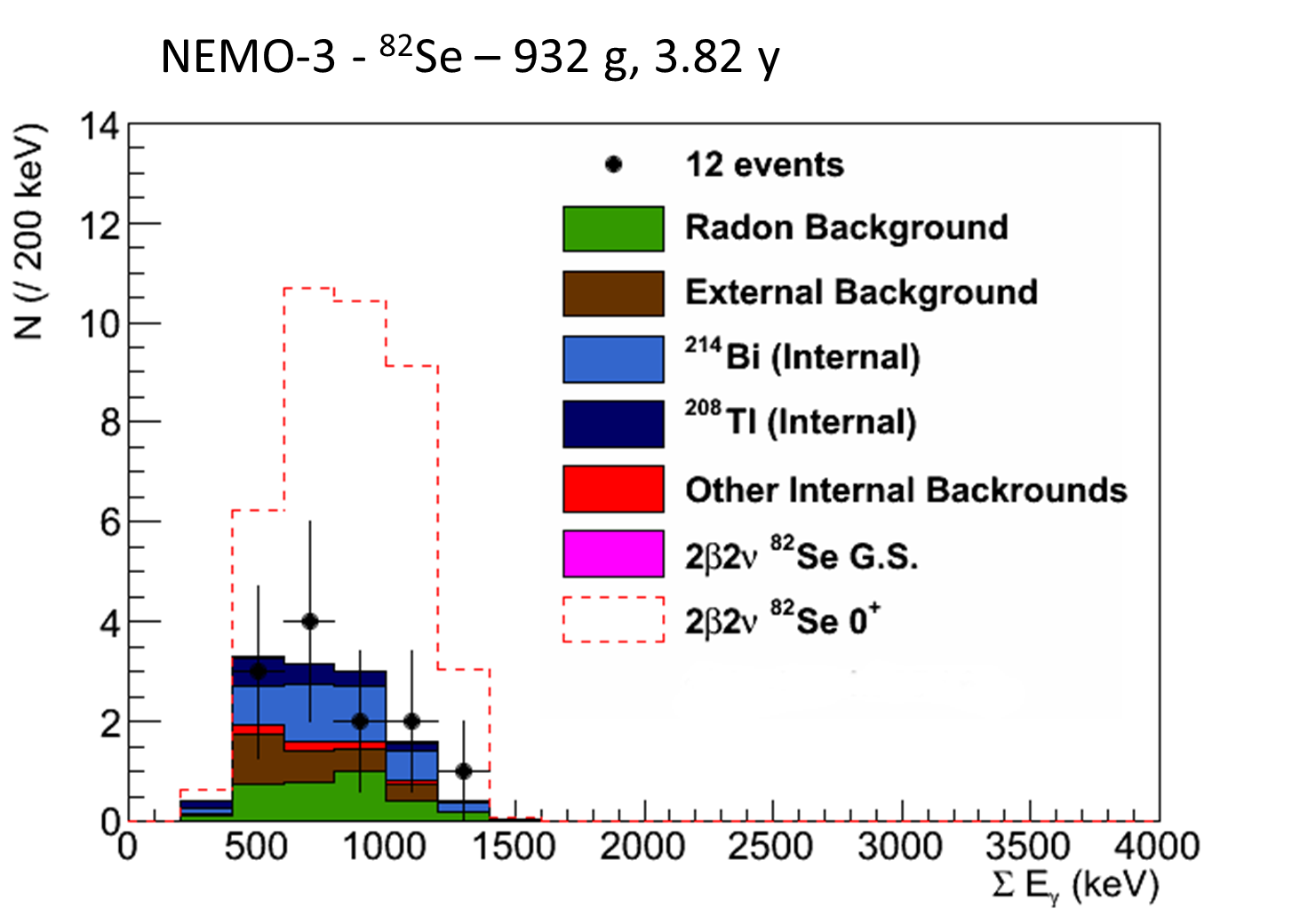} \label{fig:Eg_fin_Ph2_2n}}\\
  \caption{Total $\gamma$-rays energy distributions after selection for the $2\nu\beta\beta$($0_{gs}^{+}\rightarrow0_{1}^{+}$) transition, for Phase 1 in Figure 13(a)  and Phase 2 in Figure 13(b). Experimental data events are compared to the MC simulation for the different backgrounds. The dotted red line represents the simulated signal with a half-life of $3\times10^{20}$ years.}
  \label{fig:Eg_2n}
\end{figure}

The total electron energy distributions for Phase 1 and Phase 2 can be seen in Figure \ref{fig:Ee_2n} while the total $\gamma$-rays energy distributions are shown in Figure \ref{fig:Eg_2n}. These figures also show the different background contributions that are detailed in Table \ref{tab:expected-bckg-2n}. The largest contribution (52\% in Phase 2) comes from internal contamination of the source foils and especially from $^{214}$Bi. Radon is also responsible for 68\% of background events during Phase 1 and was reduced to 28\% in Phase 2. The external backgrounds account for 21\% of the total expected background despite the strong criteria used to ensure that only internal events are selected.

\begin{table}
\begin{centering}
\medskip{}
\begin{tabular}{|c|c|c|c|c|c|}
\cline{3-6} 
\multicolumn{2}{c|}{\selectlanguage{english}%
} & \multicolumn{2}{c|}{Expected} & \multicolumn{2}{c|}{Contribution to}\tabularnewline
\multicolumn{2}{c|}{\selectlanguage{english}%

} & \multicolumn{2}{c|}{events} & \multicolumn{2}{c|}{total background (\%)}\tabularnewline
\cline{3-6} 
\multicolumn{2}{c|}{\selectlanguage{english}%

} & Phase 1 & Phase 2 & Phase 1 & Phase 2\tabularnewline
\hline 
\multirow{4}{*}{Internal} & $^{214}$Bi & $1.14\pm0.05\pm0.12$ & $4.28\pm0.09\pm0.43$ & $12.1$ & $36.1$\tabularnewline
 & $^{208}$Tl & $0.43\pm0.02\pm0.07$ & $1.58\pm0.04\pm0.23$ & $4.6$ & $13.3$\tabularnewline
 & Others & $0.06\pm0.03\pm0.01$ & $0.29\pm0.14\pm0.02$ & $0.6$ & $2.4$\tabularnewline
\cline{2-6} 
 & Total & $1.64\pm0.07\pm0.20$ & $6.15\pm0.49\pm0.68$ & \textbf{$17.3$} & \textbf{$51.8$}\tabularnewline
\hline 
\hline 
\multicolumn{2}{|c|}{Radon} & $6.44\pm0.63\pm0.65$ & $3.26\pm0.31\pm0.33$ & \textbf{$68.0$} & \textbf{$27.5$}\tabularnewline
\hline 
\hline 
\multirow{4}{*}{External} & $^{214}$Bi & $0.38\pm0.19\pm0.04$ & $1.49\pm0.75\pm0.15$ & $4.0$ & $12.5$\tabularnewline
 & $^{208}$Tl & $0.29\pm0.10\pm0.03$ & $0.18\pm0.06\pm0.02$ & $2.9$ & $1.6$\tabularnewline
 & Others & $0.74\pm0.37\pm0.08$ & $0.78\pm0.39\pm0.08$ & $7.8$ & $6.6$\tabularnewline
\cline{2-6} 
 & Total & $1.39\pm0.43\pm0.15$ & $2.46\pm0.05\pm0.85$ & \textbf{$14.7$} & \textbf{$20.7$}\tabularnewline
\hline
\hline 
\multicolumn{2}{|c|}{Total background} & $9.47\pm0.77\pm1.00$ & $11.87\pm1.17\pm1.26$ & \textbf{$100.0$} & \textbf{$100.0$}\tabularnewline
\hline
\multicolumn{2}{|c|}{Data events} & $7$ & $12$ & \textbf{$-$} & \textbf{$-$}\tabularnewline
\hline
\end{tabular}
\par\end{centering}

\caption{Numbers of expected background events from the main background sources
in both Phases and their contribution to the total number of expected
background events for the 2e2$\gamma$
channel after optimisation for the study of $2\nu\beta\beta$($0_{gs}^{+}\rightarrow0_{1}^{+}$)
decay. 0.93 year of data taking are considered for Phase 1 and 3.82
years for Phase 2. The quoted uncertainties represent the statistical and
systematic uncertainties, respectively. The number of selected data events for each phase is also presented.}\label{tab:expected-bckg-2n}

\medskip{}
\end{table}

It is also shown that there is a good compatibility with background and data events. In the absence of a significant excess of data versus background, a limit has been
set. This can be performed using the following equation :

\begin{equation}
 T_{1/2}>\epsilon\times N_{nuc}\times\text{ln}(2)\times(t_{acq}-t_{d})\times\frac{1}{N_{ex}}
\label{eq:T1/2-simple},
\end{equation}

\noindent where $\epsilon$
  is the detection efficiency, $N_{nuc}$
  the number of $\mathrm{^{82}Se}$ nuclei, $t_{acq}$ and $t_{d}$ the acquisition and dead time respectively
  and $N_{ex}$
  the number of signal events that can be excluded.
The method used here to obtain this last number is the CLs method \cite{CLs},  that takes into account the shape of the expected signal and backgrounds as well as the number of data events and several statistical and systematical uncertainties. The systematics are detailed in Table \ref{tab:ZeroNuSystematics}. Considering then the 4.42 kg.y exposure, the 0.069\% efficiency, the 21.4 expected background events and 19 data events, the limit on the $2\nu\beta\beta$
 ($0_{gs}^{+}\rightarrow0_{1}^{+}$) decay half-life for $\mathrm{^{82}Se}$ is, at 90\% CL :

\begin{equation}
 T_{1/2}^{2\nu}({}\mathrm{^{82}Se},0_{gs}^{+}\rightarrow0_{1}^{+})>1.3\times10^{21}\:\mathrm{y.}
\end{equation}

This result is compatible with limit of $3\times10^{21}\:\mathrm{y}$
from Ref. \cite{Suhonen1997} and lower than the value published by
the LUCIFER collaboration, who determined a limit of
$3.4\times10^{22}\:\mathrm{y}$ \cite{LUCIFER-results} for the (2$\nu$+0$\nu$)$\beta\beta$ processes.
However, the NEMO-3 technique
precisely identifies the event topology and could thus independently study $2\nu\beta\beta$ and $0\nu\beta\beta$ processes.

\begin{table}[htpb]
\begin{center}
  \begin{tabular}{|l|c|c|}
\hline
    \multirow{2}{*}{Systematic} & Estimated uncertainty & \multirow{2}{*}{Method of estimate}\\
                                & (\%)            & \\
    \hline
    Gamma tracking  & 10.4  & $^{232}$U vs HPGe\\
    Energy calibration  & 1  & Neutron sources\\
    $2\nu\beta\beta$ efficiency  & 5  & $^{207}$Bi vs HPGe\\
    $^{82}$Se mass  & 0.5  & Uncertainty on mass and enrichment\\
    Energy loss in foil    & 1 & Neutron sources\\
    bremsstrahlung    & 1 & $^{90}$Y source analysis\\
    Ext. BG activities           & 10 & Variation from background model\\
    Radon BG activities          & 10 & 1e1$\alpha$ vs 1e1$\gamma$ \\
    Int. BG activities           & \multirow{2}{*}{4} & $^{207}$Bi 1eN$\gamma$ vs 2e \\
    (excl. $^{208}$Tl \& $^{214}$Bi) &                    & ($^{40}$K \& $^{234\text{m}}$Pa meas. in 1e)\\
    Int. $^{214}$Bi activity       & 10 & 1e1$\alpha$ vs. 1e1$\gamma$ \\
    Int. $^{208}$Tl activity       & 15 & NEMO-3 vs HPGe \\
    2$\nu\beta\beta$ activity    & 1  & Statistical uncertainty\\
\hline
  \end{tabular}
  \caption{Values of the $1\sigma$ systematic uncertainties included in the calculation of the limits on $2\nu\beta\beta$ decay to excited states and their methods of estimate. The estimated uncertainties come from the comparison of the activity measurements of calibration sources between NEMO-3 and HPGe ($^{232}$U, $^{207}$Bi, $^{90}$Y), the uncertainties on background measurements and uncertainties specific to the detector or $^{82}$Se sources.}
  \label{tab:ZeroNuSystematics}
\end{center}
\end{table}

\subsection{Neutrinoless double beta decay to $0_{1}^{+}$ excited state}

The search for $0\nu\beta\beta$ events is carried out similarly to what
has been done for the $2\nu\beta\beta$ decay. The preselection criteria
are the same as what is described in the first part of Section
\ref{sec:2n2b}. However, in the $0\nu\beta\beta$ process through the
$0_{gs}^{+}\rightarrow0_{1}^{+}$ transition, the two electrons do not
share energy with neutrinos contrary to the $2\nu\beta\beta$ decay. The
signal efficiency using these criteria increases by a factor 10
compared to the $2\nu\beta\beta$ process. It reaches $0.71 \%$, as higher
energy electrons are expected.  The selection has then been optimized
with these energies, taking into account a simulated signal with a
half-life of $3\times10^{21}$ years and using the same method as the
one described in Section \ref{sec:2n2b}. Applying those criteria, the
final selection efficiency for this signal is 0.69\% and 14 data events are selected.

\begin{figure}[htpb]
  \subfloat[Phase 1]{\includegraphics[width=0.5\textwidth]{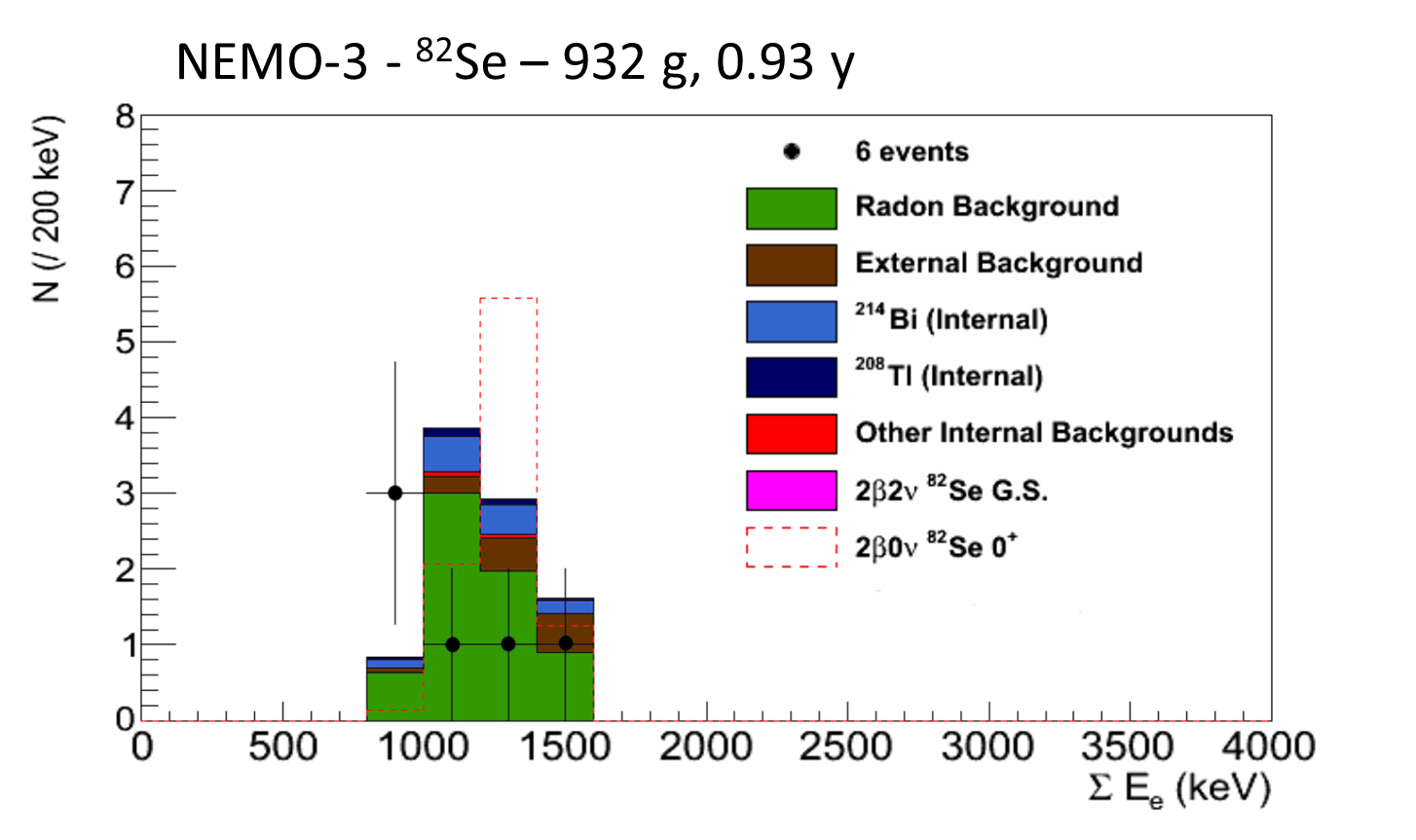} \label{fig:Ee_fin_Ph1_0n}}\hfill%
  \subfloat[Phase 2] {\includegraphics[width=0.5\textwidth]{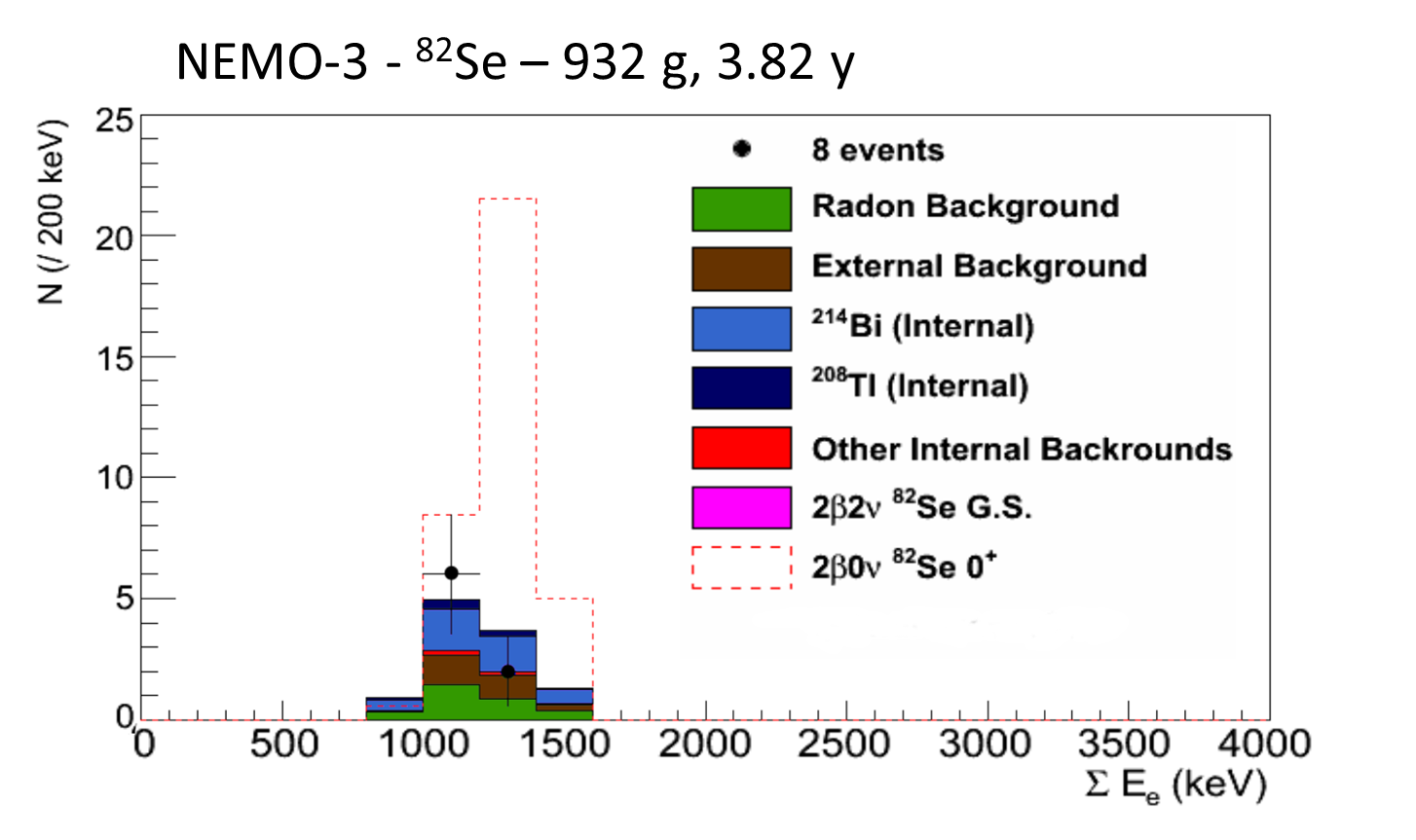} \label{fig:Ee_fin_Ph2_0n}}\\
  \caption{Total electron energy distributions after selection for the $0\nu\beta\beta$
 ($0_{gs}^{+}\rightarrow0_{1}^{+}$) transition, for Phase 1 in Figure 14(a) and Phase 2 in Figure 14(b). Experimental data events are compared to the MC simulation for the different backgrounds. The dotted red line represents the simulated signal with a half-life of $3\times10^{21}$ years.}
  \label{fig:Ee_0n}
\end{figure}

\begin{figure}[htpb]
  \subfloat[Phase 1]{\includegraphics[width=0.5\textwidth]{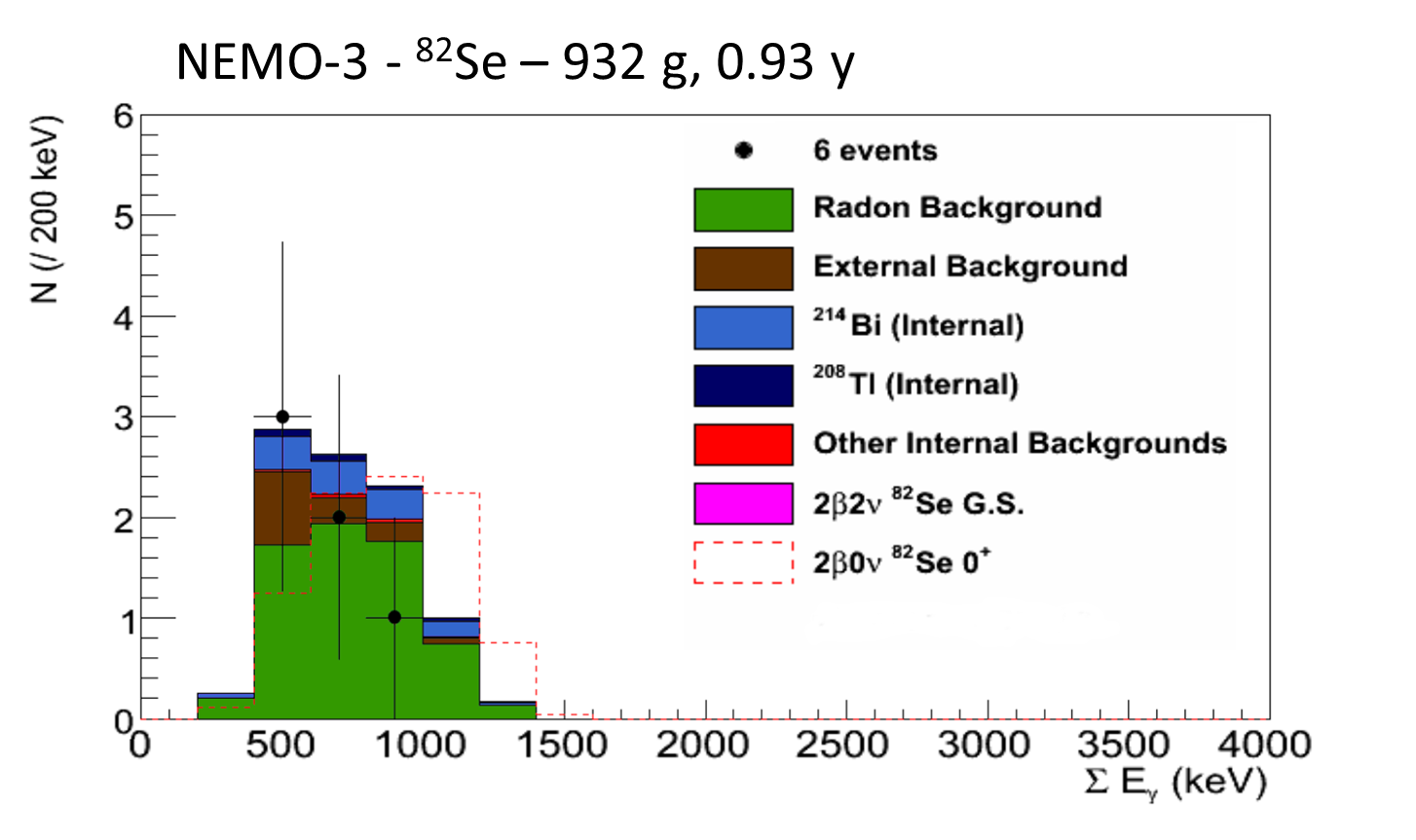} \label{fig:Eg_fin_Ph1_0n}}\hfill%
  \subfloat[Phase 2] {\includegraphics[width=0.5\textwidth]{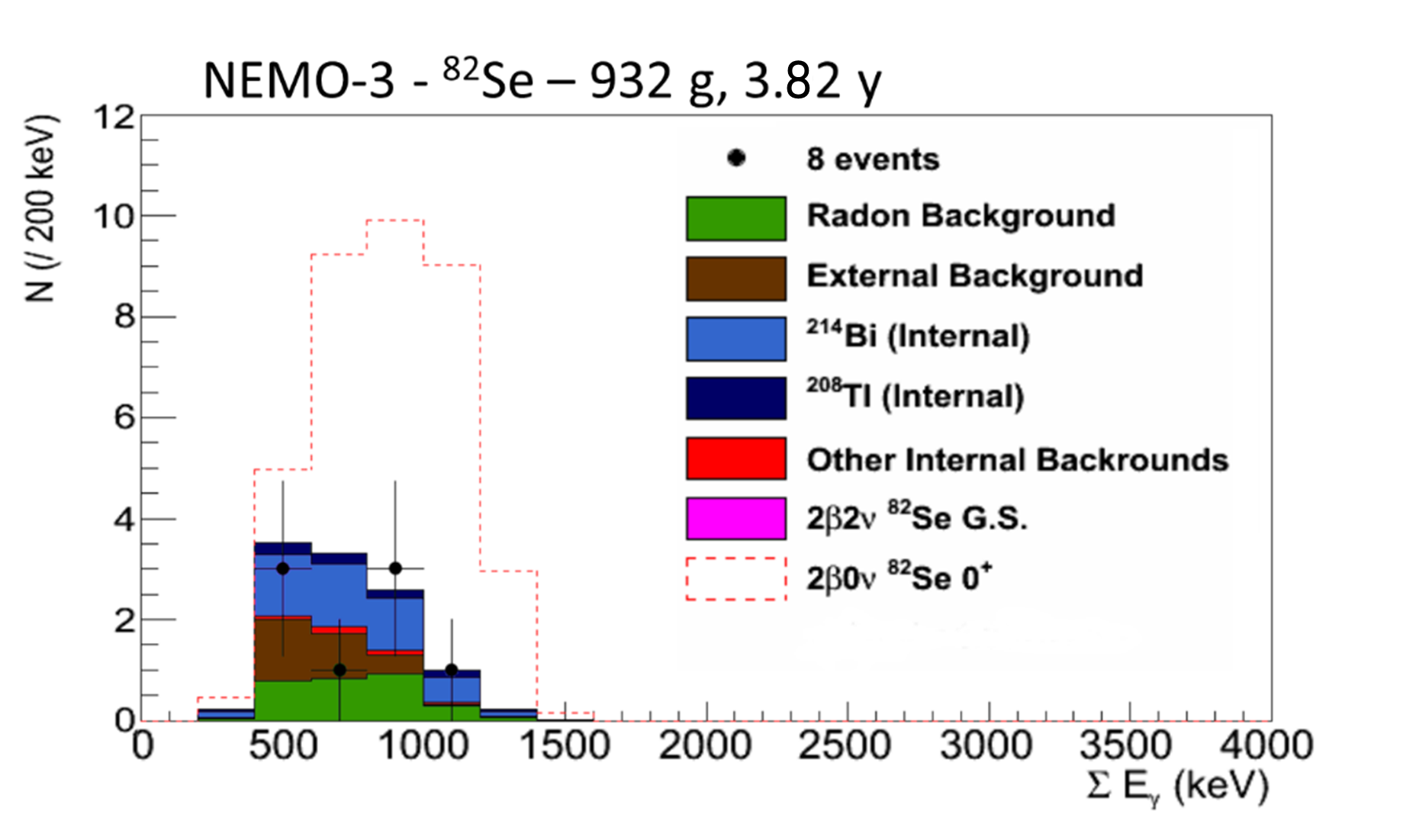} \label{fig:Eg_fin_Ph2_0n}}\\
  \caption{Total $\gamma$-rays energy distributions after selection for the $0\nu\beta\beta$
 ($0_{gs}^{+}\rightarrow0_{1}^{+}$) transition, for Phase 1 in Figure 15(a)  and Phase 2 in Figure 15(b). Experimental data events are compared to the MC simulation for the different backgrounds. The dotted red line represents the simulated signal with a half-life of $3\times10^{21}$ years.}
  \label{fig:Eg_0n}
\end{figure}

The total electron energy distributions for Phase 1 and Phase 2 are
shown in Figure \ref{fig:Ee_0n}. The background composition is similar to
what was presented in Table \ref{tab:expected-bckg-2n}  with a high radon contribution. The details are presented in Table \ref{tab:expected-bckg-0n}. The total $\gamma$-rays energy distributions are shown in Figure \ref{fig:Eg_0n}.

\begin{table}
\begin{centering}
\medskip{}
\begin{tabular}{|c|c|c|c|c|c|}
\cline{3-6} 
\multicolumn{2}{c|}{\selectlanguage{english}%
%\selectlanguage{french}%
} & \multicolumn{2}{c|}{Expected} & \multicolumn{2}{c|}{Contribution to}\tabularnewline
\multicolumn{2}{c|}{\selectlanguage{english}%
%\selectlanguage{french}%
} & \multicolumn{2}{c|}{events} & \multicolumn{2}{c|}{total background (\%)}\tabularnewline
\cline{3-6} 
\multicolumn{2}{c|}{\selectlanguage{english}%
%\selectlanguage{french}%
} & Phase 1 & Phase 2 & Phase 1 & Phase 2\tabularnewline
\hline 
\multirow{4}{*}{$^{82}$Se foils} & $^{214}$Bi & $1.25\pm0.05\pm0.13$ & $4.57\pm0.10\pm0.46$ & $13.5$ & $41.9$\tabularnewline
 & $^{208}$Tl & $0.24\pm0.01\pm0.03$ & $0.82\pm0.02\pm0.12$ & $2.6$ & $7.6$\tabularnewline
 & Others & $0.02\pm0.01\pm0.01$ & $0.07\pm0.03\pm0.01$ & $0.2$ & $0.6$\tabularnewline
\cline{2-6} 
 & Total & $1.50\pm0.06\pm0.17$ & $5.46\pm0.11\pm0.59$ & \textbf{$16.3$} & \textbf{$50.1$}\tabularnewline
\hline 
\hline 
\multicolumn{2}{|c|}{Radon} & $6.50\pm0.64\pm0.65$ & $2.97\pm0.27\pm0.30$ & \textbf{$70.4$} & \textbf{$27.3$}\tabularnewline
\hline 
\hline 
\multirow{4}{*}{Detector} & $^{214}$Bi & $0.45\pm0.22\pm0.05$ & $1.33\pm0.62\pm0.14$ & $4.9$ & $12.3$\tabularnewline
 & $^{208}$Tl & $0.79\pm0.09\pm0.08$ & $1.13\pm0.14\pm0.12$ & $8.5$ & $10.3$\tabularnewline
 & Others & $0$ & $0$ & $0$ & $0$\tabularnewline
\cline{2-6} 
 & Total & $1.24\pm0.24\pm0.13$ & $2.46\pm0.62\pm0.26$ & \textbf{$13.4$} & \textbf{$22.6$}\tabularnewline
\hline
\hline 
\multicolumn{2}{|c|}{Total background} & $9.24\pm0.69\pm0.95$ & $10.89\pm0.69\pm1.15$ & \textbf{$100.0$} & \textbf{$100.0$}\tabularnewline
\hline
\multicolumn{2}{|c|}{Data events} & $6$ & $9$ & \textbf{$-$} & \textbf{$-$}\tabularnewline
\hline
\end{tabular}
\par\end{centering}

\caption{Numbers of expected background events from the main background sources
in both Phases and their contribution to the total number of expected
background events for the 2e2$\gamma$
channel after optimisation for the study of $0\nu\beta\beta$($0_{1}^{+}\rightarrow0_{2}^{+}$)
decay. 0.93 years of data taking are considered for Phase 1 and 3.82
years for Phase 2. The quoted uncertainties represent the statistical and
systematic uncertainties, respectively. The number of selected data events for each phase is also presented.}\label{tab:expected-bckg-0n}

\medskip{}
\end{table}

As for the $2\nu\beta\beta$($0_{gs}^{+}\rightarrow0_{1}^{+}$) transition,
data is consistent with background-only predictions so a limit has to be set on the half-life of the $0\nu\beta\beta$($0_{gs}^{+}\rightarrow0_{1}^{+}$) process. The method used to calculate such a limit remains the CLs method. With the 20.1 background events and the statistical and systematic uncertainties, the limit on the $0\nu\beta\beta$
 ($0_{gs}^{+}\rightarrow0_{1}^{+}$) decay half-life for $^{82}$Se (at 90\% CL) is :

\begin{equation}
 T_{1/2}^{0\nu}({}^{82}\mathrm{Se},0_{gs}^{+}\rightarrow0_{1}^{+})>2.3\times10^{22}\:\mathrm{y}.
\end{equation}

This result is  given for the
$0\nu\beta\beta$($0_{gs}^{+}\rightarrow0_{1}^{+}$) transition for
$^{82}$Se, separately from
$2\nu\beta\beta$($0_{gs}^{+}\rightarrow0_{1}^{+}$). It is compatible with
limit of 3.4$\times10^{22}\:~\mathrm{y}$ and
8.1$\times10^{22}\:~\mathrm{y}$ obtained in the LUCIFER experiment \cite{LUCIFER-results}
 for the (2$\nu$+0$\nu$)$\beta\beta$ processes and CUPID-0 \cite{CUPID-0} experiment. According to the mass
mechanism, a Majorana neutrino is exchanged during such a process and therefore a limit can also be set on the effective mass of the neutrino using the following equation :

\begin{equation}
\frac{1}{\left(T_{1/2}^{0\nu}\right)_{MM}}=G^{0\nu}(Q_{\beta\beta},Z)
g_A^4 \left|M^{0\nu}\right|^{2}\left|\frac{m_{\beta\beta}}{m_{e}}\right|^{2},
\end{equation}

\noindent where $G^{0\nu}(Q_{\beta\beta},Z)$
  is the phase space factor given in \cite{EspaceDePhaseStoica} for
  the transition, $\mathrm{g_A = 1.27}$ and $M^{0\nu}$
  the nuclear matrix element \cite{Simkovic, Menendez2009,Hyvarinen2016, Barea2015}. The limit that can be set on the effective neutrino mass is $m_{\beta\beta}<\left[42-239\right]\:\text{eV}$.

\section{Summary and Conclusions}
4
Using an innovative gamma tracking technique, the NEMO-3 data set was
analysed to search for $\beta\beta$ decays of $^{82}$Se to the excited
states of $^{82}$Kr with a 4.42 kg.y exposure. No evidence for the
$2\nu\beta\beta$ process was found and thus an upper limit on the
decay half-life was set at 90\% CL :
$T_{1/2}^{2\nu}({}^{82}\mathrm{Se},0_{gs}^{+}\rightarrow0_{1}^{+})>1.3\times10^{21}\:\mathrm{y}$.
This result can nevertheless help to constrain theoretical QRPA models
presented in \cite{Suhonen,QRPAWS,QRPA2nu}.

The analysis of the $0\nu\beta\beta$ decay to excited states was
conducted in a similar fashion and, as once again no extra events were
observed over the expected background, an upper limit was set at 90\%
CL :
$T_{1/2}^{0\nu}({}^{82}\mathrm{Se},0_{gs}^{+}\rightarrow0_{1}^{+})>2.3\times10^{22}\:\mathrm{y}$.
These results are obtained for the first
time with a detector which reconstructs each particle individually in the final state.

This analysis performed with $^{82}$Se in NEMO-3 will also provide useful information for the next-generation SuperNEMO experiment which will host 100 kg of $^{82}$Se, such as optimisation of the selected events and identification of the main background contributions. 

In parallel with its search for $0\nu\beta\beta$ decay to the ground
state, SuperNEMO will also look for the $2\nu\beta\beta$ and
$0\nu\beta\beta$ decays to excited states with major
improvements. Using thicker scintillators, the sensitivity to $\gamma$-rays and efficiency to $2\nu\beta\beta$ and
$0\nu\beta\beta$($0_{gs}^{+}\rightarrow0_{1}^{+}$) transitions will be
enhanced. Backgrounds will also be reduced : more than a factor 30 for
radon and a factor 100 for $^{214}$Bi and $^{208}$Tl. The expected
sensitivities for SuperNEMO are respectively $\sim$  $10^{23}$ y and
$\sim$ $10^{24}$ y for the $2\nu\beta\beta$ and
$0\nu\beta\beta$($0_{gs}^{+}\rightarrow0_{1}^{+}$) half-lives. A first
module, called Demonstrator, with 7 kg of
$^{82}$Se is undergoing commissioning and will start
taking data in 2019. Its goal is to reach a sensitivity on the
$0\nu\beta\beta$ half-life of $5\times10^{24}\:\mathrm{y}$ in 17.5
kg$^.$y exposure with the demonstration of a ``zero"-background
experiment \cite{Fred_ICHEP2016}.

{\bf Acknowledgements} We thank the staff of the Modane Underground Laboratory for their technical assistance in running the experiment. We acknowledge support by the grants agencies of the Czech Republic (grant number EF16 013/0001733), CNRS/IN2P3 in France, RFBR  in Russia (Project No.19-52-16002 NCNIL a), APVV in Slovakia (Project No. 15-0576), STFC in the UK and NSF in the USA. 

%\section*{References}

\bibliography{NEMO3_Se82}

\begin{thebibliography}{25}
\expandafter\ifx\csname natexlab\endcsname\relax\def\natexlab#1{#1}\fi
\providecommand{\url}[1]{\texttt{#1}}
\providecommand{\href}[2]{#2}
\providecommand{\path}[1]{#1}
\providecommand{\DOIprefix}{doi:}
\providecommand{\ArXivprefix}{arXiv:}
\providecommand{\URLprefix}{URL: }
\providecommand{\Pubmedprefix}{pmid:}
\providecommand{\doi}[1]{\href{http://dx.doi.org/#1}{\path{#1}}}
\providecommand{\Pubmed}[1]{\href{pmid:#1}{\path{#1}}}
\providecommand{\bibinfo}[2]{#2}
\ifx\xfnm\relax \def\xfnm[#1]{\unskip,\space#1}\fi
%Type = Article
\bibitem[{et~al. (NEMO-3~Collaboration)(2015)}]{NEMO32015}
\bibinfo{author}{R.~Arnold et~al. (NEMO-3~Collaboration)}
  (\bibinfo{collaboration}{NEMO-3 Collaboration}), \bibinfo{journal}{Phys. Rev.
  D} \bibinfo{volume}{\textbf{92}} (\bibinfo{year}{2015}) \bibinfo{pages}{072011}.
  \URLprefix \url{https://link.aps.org/doi/10.1103/PhysRevD.92.072011}.
  \DOIprefix\doi{10.1103/PhysRevD.92.072011}.
%Type = Article
%\bibitem[{et~al. (CUORE~Collaboration)(2016)}]{CUORE2016}
%\bibinfo{author}{C.~Alduino et~al. (CUORE~Collaboration)}
%  (\bibinfo{collaboration}{CUORE Collaboration}), \bibinfo{journal}{Phys. Rev.
%  C} \bibinfo{volume}{93} (\bibinfo{year}{2016}) \bibinfo{pages}{045503}.
%  \URLprefix \url{https://link.aps.org/doi/10.1103/PhysRevC.93.045503}.
%  \DOIprefix\doi{10.1103/PhysRevC.93.045503}.
%Type = Article
\bibitem[{et~al. (CUORE~Collaboration)(2018)}]{CUORE2018}
\bibinfo{author}{C.~Alduino et~al. (CUORE~Collaboration)}
  (\bibinfo{collaboration}{CUORE Collaboration}), \bibinfo{journal}{Phys. Rev.
  Lett.} \bibinfo{volume}{\textbf{120}} (\bibinfo{year}{2018}) \bibinfo{pages}{132501}.
  \URLprefix \url{https://link.aps.org/doi/10.1103/PhysRevLett.120.132501}.
  \DOIprefix\doi{10.1103/PhysRevLett.120.132501}.
%Type = Article
%\bibitem[{et~al. (GERDA~Collaboration)(2013)}]{GERDA2013}
%\bibinfo{author}{M.~Agostini et~al. (GERDA~Collaboration)}
%  (\bibinfo{collaboration}{GERDA Collaboration}), \bibinfo{journal}{Phys. Rev.
%  Lett.} \bibinfo{volume}{111} (\bibinfo{year}{2013}) \bibinfo{pages}{122503}.
%  \URLprefix \url{https://link.aps.org/doi/10.1103/PhysRevLett.111.122503}.
%  \DOIprefix\doi{10.1103/PhysRevLett.111.122503}.
%Type = Article
\bibitem[{et~al. (GERDA~Collaboration)(2018)}]{GERDA2018}
\bibinfo{author}{M.~Agostini et~al. (GERDA~Collaboration)}
  (\bibinfo{collaboration}{GERDA Collaboration}), \bibinfo{journal}{Phys. Rev.
  Lett.} \bibinfo{volume}{\textbf{120}} (\bibinfo{year}{2018}) \bibinfo{pages}{132503}.
  \URLprefix \url{https://link.aps.org/doi/10.1103/PhysRevLett.120.132503}.
  \DOIprefix\doi{10.1103/PhysRevLett.120.132503}.
%Type = Article
\bibitem[{et~al. (KamLAND-Zen~Collaboration)(2016)}]{Gando2016}
\bibinfo{author}{A.~Gando et~al. (KamLAND-Zen~Collaboration)}
  (\bibinfo{collaboration}{KamLAND-Zen Collaboration}), \bibinfo{journal}{Phys.
  Rev. Lett.} \bibinfo{volume}{\textbf{117}} (\bibinfo{year}{2016})
  \bibinfo{pages}{082503}. \URLprefix
  \url{https://link.aps.org/doi/10.1103/PhysRevLett.117.082503}.
  \DOIprefix\doi{10.1103/PhysRevLett.117.082503}.
%Type = Article
\bibitem[{et~al.(2013)}]{Lincoln2013}
\bibinfo{author}{D.~L.~Lincoln et~al.}, \bibinfo{journal}{Phys. Rev. Lett.}
  \bibinfo{volume}{\textbf{110}} (\bibinfo{year}{2013}) \bibinfo{pages}{012501}.
%Type = Article
\bibitem[{J.~W.~Beeman et~al. (LUCIFER~Collaboration)(2013)}]{LUCIFER}
\bibinfo{author}{J.~W.~Beeman et~al. (LUCIFER~Collaboration)},
  \bibinfo{journal}{Adv. High Energy Phys.} \bibinfo{volume}{\textbf{2013}}
  (\bibinfo{year}{2013}) \bibinfo{pages}{237973}.
%Type = Article
%\bibitem[{Bellini et~al.(2019)}]{CUPID-0_v2}
%\bibinfo{author}{F.~Bellini}, et~al., \bibinfo{journal}{Universe} \bibinfo{volume}{\textbf{5}}
%  (\bibinfo{year}{2019}) \bibinfo{pages}{2} \DOIprefix\doi{10.3390/universe5010002}.
\bibitem[{O.Azzolini et~al.(2018)}]{CUPID-0_v2-detector}
\bibinfo{author}{O.~Azzolini}, et~al.,
\bibinfo{journal}{Eur. Phys. J. C} \bibinfo{volume}{\textbf{78}}
  (\bibinfo{year}{2018}) \bibinfo{pages}{428}
%Type = Article
\bibitem[{et~al.(2010)}]{SuperNEMO}
\bibinfo{author}{R.~Arnold et~al.}, \bibinfo{journal}{Eur. Phys. J. C}
  \bibinfo{volume}{\textbf{70}} (\bibinfo{year}{2010})
  \bibinfo{pages}{927--943}.
%Type = Article
\bibitem[{Arnold et~al.(2018)}]{Arnold:2018se}
\bibinfo{author}{R.~Arnold}, et~al. (\bibinfo{collaboration}{NEMO-3
  Collaboration}), \bibinfo{journal}{Eur. Phys. J. C} \bibinfo{volume}{\textbf{78}}
  (\bibinfo{year}{2018}) \bibinfo{pages}{821}.
%Type = Article
\bibitem[{O.Azzolini et~al.(2019)}]{CUPID-0_v2}
\bibinfo{author}{O.~Azzolini}, et~al., \bibinfo{journal}{Phys. Rev. Lett.} \bibinfo{volume}{\textbf{123}}
  (\bibinfo{year}{2019}) \bibinfo{pages}{032501} \DOIprefix\doi{10.3390/PhysRevLett.123.032501}.
%Type = Article
\bibitem[{Menéndez et~al.(2009)Menéndez, Poves, Caurier, and
  Nowacki}]{MENENDEZ2009139}
\bibinfo{author}{J.~Men\'endez}, \bibinfo{author}{A.~Poves},
  \bibinfo{author}{E.~Caurier}, \bibinfo{author}{F.~Nowacki},
  \bibinfo{journal}{Nuclear Physics A} \bibinfo{volume}{\textbf{818}}
  (\bibinfo{year}{2009}) \bibinfo{pages}{139--151}. \URLprefix
  \url{http://www.sciencedirect.com/science/article/pii/S0375947408008233}.
  \DOIprefix\doi{http://dx.doi.org/10.1016/j.nuclphysa.2008.12.005}.
%Type = Article
\bibitem[{Barea et~al.(2015)Barea, Kotila, and Iachello}]{Barea2015}
\bibinfo{author}{J.~Barea}, \bibinfo{author}{J.~Kotila},
  \bibinfo{author}{F.~Iachello}, \bibinfo{journal}{Phys. Rev. C}
  \bibinfo{volume}{\textbf{91}} (\bibinfo{year}{2015}) \bibinfo{pages}{034304}.
  \URLprefix \url{https://link.aps.org/doi/10.1103/PhysRevC.91.034304}.
  \DOIprefix\doi{10.1103/PhysRevC.91.034304}.
%Type = Article
\bibitem[{Hyv\"arinen and Suhonen(2016)}]{Hyvarinen2016}
\bibinfo{author}{J.~Hyv\"arinen}, \bibinfo{author}{J.~Suhonen},
  \bibinfo{journal}{Phys. Rev. C} \bibinfo{volume}{\textbf{93}} (\bibinfo{year}{2016})
  \bibinfo{pages}{064306}. \URLprefix
  \url{https://link.aps.org/doi/10.1103/PhysRevC.93.064306}.
  \DOIprefix\doi{10.1103/PhysRevC.93.064306}.
%Type = Article
\bibitem[{Dolinski(2019)}]{Dolinski2019}
\bibinfo{author}{M. J. Dolinski et al},
  \bibinfo{journal}{Annu. Rev. Nucl. Part. Sci.} \bibinfo{volume}{\textbf{69}} (\bibinfo{year}{2019})
  \bibinfo{pages}{219-251}. \URLprefix
  \url{https://www.annualreviews.org/doi/10.1146/annurev-nucl-101918-023407}.
  \DOIprefix\doi{10.1146/annurev-nucl-101918-023407}.
%Type = Article
\bibitem[{Barabash(2015)}]{Barabash2015}
\bibinfo{author}{A.~S.~Barabash},
  \bibinfo{journal}{Nucl. Phys. A} \bibinfo{volume}{\textbf{393}} (\bibinfo{year}{2015})
  \bibinfo{pages}{593}. \URLprefix
  \url{https://www.sciencedirect.com/science/article/pii/S037594741500010X?via%3Dihub}.
  \DOIprefix\doi{10.1016/j.nuclphysa.2015.01.001}.
%Type = Article
\bibitem[{Barabash(2019)}]{Barabash2019}
\bibinfo{author}{A.~S.~Barabash},
  \href{https://arxiv.org/abs/1907.06887}{\tt arXiv:nucl-ex/190706887}.
%Type = Article
\bibitem[{PRL(2005)}]{NEMO3:FirstResult}
\bibinfo{author}{R.~Arnold} et~al. (\bibinfo{collaboration}{NEMO-3
  Collaboration}),
  \bibinfo{journal}{Phys. Rev. Lett.} \bibinfo{volume}{\textbf{95}} (\bibinfo{year}{2005})
  \bibinfo{pages}{182302}. \URLprefix
\url{https://link.aps.org/doi/10.1103/PhysRevLett.120.132501}.
  \DOIprefix\doi{10.1103/PhysRevLett.120.132501}.  
%Type = Article
\bibitem[{et~al.(2014)}]{MoExciteHPGe}
\bibinfo{author}{R.~Arnold et~al.}, \bibinfo{journal}{Nucl. Phys. A}
  \bibinfo{volume}{\textbf{925}} (\bibinfo{year}{2014}) \bibinfo{pages}{25--36}.
%Type = Article
\bibitem[{et~al.(2007)}]{excitedMo}
\bibinfo{author}{R.~Arnold et~al.}, \bibinfo{journal}{Nucl. Phys. A}
  \bibinfo{volume}{\textbf{781}} (\bibinfo{year}{2007}) \bibinfo{pages}{209--226}.
%Type = Article
\bibitem[{Arnold(2016)}]{NEMO3:150Nd}
\bibinfo{author}{R.~Arnold}, et~al. (\bibinfo{collaboration}{NEMO-3
  Collaboration}), \bibinfo{journal}{Phys. Rev. D} \bibinfo{volume}{\textbf{94}}
  (\bibinfo{year}{2016}) \bibinfo{pages}{072003}.
  \url{https://link.aps.org/doi/10.1103/PhysRevD.94.072003}.
  \DOIprefix\doi{10.1103/PhysRevD.94.072003}.
%Type = Article
\bibitem[{A.~S.~Barabash and Umatov(2009)}]{NdExciteHPGe}
\bibinfo{author}{A.~S.~Barabash, Ph.~Hubert, A.~Nachab and
  V.~I.~Umatov},
 \bibinfo{journal}{Phys. Rev. C} \bibinfo{volume}{\textbf{79}}
  (\bibinfo{year}{2009}) \bibinfo{pages}{45501}.
%Type = Article
\bibitem[{M.~F.~Kidd(2014)}]{NdExciteHPGe2}
\bibinfo{author}{M.~F.~Kidd, J. H.~Esterline, S.~W.~Finch and W.~Tornow},
  \bibinfo{journal}{Phys. Rev. C} \bibinfo{volume}{\textbf{90}}
  (\bibinfo{year}{2014}) \bibinfo{pages}{055501}.
%Type = Article
\bibitem[{Kotila and Iachello(2012)}]{Iachello2012}
\bibinfo{author}{J.~Kotila}, \bibinfo{author}{F.~Iachello},
  \bibinfo{journal}{Phys. Rev. C} \bibinfo{volume}{\textbf{85}}
  (\bibinfo{year}{2012})   \bibinfo{pages}{034316}.
%Type = Article
\bibitem[{Arnold et~al.(2016)}]{ReviewBarabash}
\bibinfo{author}{A.~S.~Barabash},  \bibinfo{journal}{AIP Conf. Proc.} \bibinfo{volume}{\textbf{1894}}
  (\bibinfo{year}{2017}) \bibinfo{pages}{020002}.
%Type = Article
\bibitem[{A.~S.~Barabash(1995)}]{Barabash1995}
\bibinfo{author}{A.~S.~Barabash},
  \bibinfo{journal}{Phys. Lett. B} \bibinfo{volume}{\textbf{345}}
  (\bibinfo{year}{1995}) \bibinfo{pages}{408--413}.
%Type = Article
\bibitem[{A.~S.~Barabash(1999)}]{Barabash1999}
\bibinfo{author}{A.~S.~Barabash},
  \bibinfo{journal}{Phys. At. Nucl.} \bibinfo{volume}{\textbf{62}}
  (\bibinfo{year}{1999}) \bibinfo{pages}{2039--2043}.
%Type = Article
\bibitem[{L.~De Braeckeleer(2001)}]{Braeckeleer}
\bibinfo{author}{L.~De Braeckeleer, M. Hornish, A.~S.~Barabash, V.~Umatov},
  \bibinfo{journal}{Phys. Rev. Lett.} \bibinfo{volume}{\textbf{86}}
  (\bibinfo{year}{2001}) \bibinfo{pages}{3510--3513}.
%Type = Article
\bibitem[{M.~F.~Kidd(2009)}]{Kidd2009}
\bibinfo{author}{M.~F.~Kidd et al},
  \bibinfo{journal}{Nucl. Phys. A} \bibinfo{volume}{\textbf{821}}
  (\bibinfo{year}{2009}) \bibinfo{pages}{251--261}.
%Type = Article
\bibitem[{P.~Belli(2010)}]{Belli}
\bibinfo{author}{P.~Belli et al},
  \bibinfo{journal}{Nucl. Phys. A} \bibinfo{volume}{\textbf{846}}
  (\bibinfo{year}{2010}) \bibinfo{pages}{143--156}.
%Type = Phdthesis
\bibitem[{Blondel(2013)}]{theseSophie}
\bibinfo{author}{S.~Blondel}, \bibinfo{title}{\textup{Optimisation du blindage
  contre les neutrons pour le d\'{e}monstrateur de {S}uper{NEMO} et analyse de
  la double d\'{e}sint\'{e}gration $\beta$ du n\'{e}odyme-150 vers les
  \'{e}tats excit\'{e}s du samarium-150 avec le d\'{e}tecteur {NEMO} 3}},
  \bibinfo{type}{PhD thesis}, Universit\'{e} Paris Sud, \bibinfo{year}{2013}.
%Type = Article
\bibitem[{J.~W.~Beeman et~al. (LUCIFER~Collaboration)(2015)}]{LUCIFER-results}
\bibinfo{author}{J.~W.~Beeman et~al. (LUCIFER~Collaboration)},
  \bibinfo{journal}{Eur. Phys. J. C} \bibinfo{volume}{\textbf{75}}
  (\bibinfo{year}{2015}) \bibinfo{pages}{591}.
%Type = Article
\bibitem[{O. Azzolinini et~al. (CUPID-0~Collaboration)(2018)}]{CUPID-0}
\bibinfo{author}{O. Azzolini et~al. (CUPID-0~Collaboration)},
  \bibinfo{journal}{Eur. Phys. J. C} \bibinfo{volume}{\textbf{78}}
  (\bibinfo{year}{2018}) \bibinfo{pages}{888}.
%Type = Article
\bibitem[{Aunola and Suhonen(1996)}]{QRPAWS}
\bibinfo{author}{M.~Aunola}, \bibinfo{author}{J.~Suhonen},
  \bibinfo{journal}{Nucl. Phys. A} \bibinfo{volume}{\textbf{602}}
  (\bibinfo{year}{1996}) \bibinfo{pages}{133--166}.
%Type = Article
\bibitem[{Suhonen and Civitarese(1998)}]{QRPA2nu}
\bibinfo{author}{J.~Suhonen}, \bibinfo{author}{O.~Civitarese},
  \bibinfo{journal}{Phys. Rep.} \bibinfo{volume}{\textbf{300}}
  (\bibinfo{year}{1998}) \bibinfo{pages}{123--214}.
%Type = Article
\bibitem[{Arnold et~al.(2005)Arnold, Augier, Bakalyarov, Baker, Barabash
  et~al.}]{Arnold:2004xq}
\bibinfo{author}{R.~Arnold}, \bibinfo{author}{C.~Augier},
  \bibinfo{author}{A.~Bakalyarov}, \bibinfo{author}{J.~Baker},
  \bibinfo{author}{A.~Barabash}, et~al.,
  \bibinfo{journal}{Nucl. Instrum. Meth.}
  \bibinfo{volume}{\textbf{A 536}} (\bibinfo{year}{2005}) \bibinfo{pages}{79--122}.
  \DOIprefix\doi{10.1016/j.nima.2004.07.194}.
  \href{http://arxiv.org/abs/physics/0402115}{\tt arXiv:physics/0402115}.
%Type = Book
%\bibitem[{Table of Isotopes(1996)}]{TableOfIsotopes}
%\bibinfo{author}{R. B.~Firestone},
%   \bibinfo{editor}{edited by
%    V. S. Shirley}, \bibinfo{title}{Table Of Isotopes}, \bibinfo{publisher}{Wiley-Interscience},
%  \bibinfo{year}{1996}.
%Type = Book
\bibitem[{Table of Isotopes(1996)}]{TableOfIsotopes}
\bibinfo{author}{J. K.~Tuli}, \bibinfo{author}{E.~Brown}, 
\bibinfo{journal}{Nucl. Data Sheets}, \bibinfo{volume}{\textbf{157}},
\bibinfo{pages}{260},
  \bibinfo{year}{2019}.
%Type = Article
\bibitem[{Suhonen et~al.(1997)Suhonen, Toivanen, Barabash, Vanushin, Umatov,
  Gurriar{\'a}n, Hubert, and Hubert}]{Suhonen1997}
\bibinfo{author}{J.~Suhonen}, \bibinfo{author}{J.~Toivanen},
  \bibinfo{author}{A.~S. Barabash}, \bibinfo{author}{I.~A. Vanushin},
  \bibinfo{author}{V.~I. Umatov}, \bibinfo{author}{R.~Gurriar{\'a}n},
  \bibinfo{author}{F.~Hubert}, \bibinfo{author}{P.~Hubert},
  \bibinfo{journal}{Zeitschrift f{\"u}r Physik A Hadrons and Nuclei}
  \bibinfo{volume}{\textbf{358}} (\bibinfo{year}{1997}) \bibinfo{pages}{297--301}.
  \URLprefix \url{http://dx.doi.org/10.1007/s002180050333}.
  \DOIprefix\doi{10.1007/s002180050333}.
%Type = Article
%\bibitem[{O. Azzolini(2018)}]{CUPID-0-excited}
%\bibinfo{author}{O.~Azzolini et al},
%  \bibinfo{journal}{Eur. Phys. J. C} \bibinfo{volume}{\textbf{78}}
%  (\bibinfo{year}{2018}) \bibinfo{pages}{888}.

%Type = Article
\bibitem[{Argyriades et~al.(2009)}]{Argyriades:2009vq}
\bibinfo{author}{J.~Argyriades}, et~al. (\bibinfo{collaboration}{NEMO-3
  Collaboration}), \bibinfo{journal}{Nucl. Instrum. Meth.}
\bibinfo{volume}{\textbf{A 606}}
  (\bibinfo{year}{2009}) \bibinfo{pages}{449--465}.
  \DOIprefix\doi{10.1016/j.nima.2009.04.011}.
  \href{http://arxiv.org/abs/0903.2277}{\tt arXiv:0903.2277}.
%Type = Article
\bibitem[{O. Ponkratenko et~al. (2000)}]{DECAY0}
\bibinfo{author}{O. Ponkratenko}, \bibinfo{author}{V. Tretyak}, \bibinfo{author}{Y. Zdesenko},
 \bibinfo{journal}{Phys. Atom. Nucl.} \bibinfo{volume}{\textbf{63}}
  (\bibinfo{year}{2000}) \bibinfo{pages}{1282}.
%Type = Article
\bibitem[{Arnold et~al.(2018)}]{GEANT3}
\bibinfo{author}{R. Brun}, \bibinfo{author}{F. Bruyant}, \bibinfo{author}{M.Maire}, \bibinfo{author}{A.McPherson}, \bibinfo{author}{P. Zanarini}
 \bibinfo{journal}{CERNDD-
EE-84-1} 
  (\bibinfo{year}{1987}) .
%Type = Phdthesis
\bibitem[{Hugon(2012)}]{theseHugon}
\bibinfo{author}{C.~Hugon}, \bibinfo{title}{\textup{Analyse des donn\'{e}es de
  l'exp\'{e}rience {NEMO}3 pour la recherche de la d\'{e}sint\'{e}gration
  double b\^{e}ta sans \'{e}mission de neutrinos. \'{E}tude des biais
  syst\'{e}matiques du calorim\`{e}tre et d\'{e}veloppement d'outils
  d'analyse}}, \bibinfo{type}{PhD thesis}, Universit\'{e} Paris Sud,
  \bibinfo{year}{2012}.
%Type = Article
\bibitem[{T.~Junk(1999)}]{CLs}
\bibinfo{author}{T.~Junk},   \bibinfo{journal}{Nucl. Instrum. Meth.}
\bibinfo{volume}{\textbf{A 434}}
  (\bibinfo{year}{1999}) \bibinfo{pages}{435--443}.
  \DOIprefix\doi{10.1016/S0168-9002(99)00498-2}.
  \href{https://arxiv.org/abs/hep-ex/9902006}{\tt arXiv:hep-ex/9902006}.
%Type = Article
%\bibitem[{Kotila and Iachello(2012)}]{EspaceDePhase}
%\bibinfo{author}{J.~Kotila}, \bibinfo{author}{F.~Iachello},
%  \bibinfo{journal}{Phys. Rev. C} \bibinfo{volume}{\textbf{85}}
%  (\bibinfo{year}{2012}).
%Type = Article
%\bibitem[{S. Stoica and M. Mirea(2013)}]{EspaceDePhaseStoica}
%\bibinfo{author}{S.~Stoica and M.~Mirea}, 
%  \bibinfo{journal}{Phys. Rev. C} \bibinfo{volume}{\textbf{88}}
%  (\bibinfo{year}{2013}).
%Type = Article
\bibitem[{S. Stoica and M. Mirea(2015)}]{EspaceDePhaseStoica}
\bibinfo{author}{M.~Mirea, T.~Pahomi, S.~Stoica}, 
  \bibinfo{journal}{Rom. Rep. Phys.} \bibinfo{volume}{\textbf{67}}
  (\bibinfo{year}{2015}) \bibinfo{pages}{872--889}.
%Type = Article
\bibitem[{F.~Simkovic and Faessler(2001)}]{Simkovic}
\bibinfo{author}{F.~\v{S}imkovic, M.~Nowak, W.~A.~Kami\'nski, A.~A.~Raduta
  and ~A.~Faessler}
  \bibinfo{journal}{Phys. Rev. C}
  \bibinfo{volume}{\textbf{64}} (\bibinfo{year}{2001}) \bibinfo{pages}{035501}.
%Type = Article
\bibitem[J. Men\'endez, A. Poves, E. Caurier, F. Nowacki (2009)]{Menendez2009}
\bibinfo{author}{J.~Men\'endez, A.~Poves, E.~Caurier, F.~Nowacki},
  \bibinfo{journal}{Nucl. Phys. A}
  \bibinfo{volume}{\textbf{818}} (\bibinfo{year}{2009}) \bibinfo{pages}{139--151}.
%Type = Article
%\bibitem[{J. Barea, J. Kotila and F. Iachello(2015)}]{Iachello2015}
%\bibinfo{author}{J.~Barea, J.~Kotila and F.~Iachello},
%  \bibinfo{journal}{Phys. Rev. C} \bibinfo{volume}{\textbf{91}}
%  (\bibinfo{year}{2015}) \bibinfo{pages}{034304}.
%Type = Article
\bibitem[{Toivanen and Suhonen(1997)}]{Suhonen}
\bibinfo{author}{J.~Toivanen}, \bibinfo{author}{J.~Suhonen},
  \bibinfo{journal}{Phys. Rev. C} \bibinfo{volume}{\textbf{55}}
  (\bibinfo{year}{1997}) \bibinfo{pages}{2314--2323}.
%Type = Proceeding
\bibitem[{Perrot(2016)}]{Fred_ICHEP2016}
\title{Status of SuperNEMO Demonstator, 38th International Conference
  on High Energy Physics (ICHEP 2016)}
\bibinfo{author}{F.~Perrot (on behalf of the SuperNEMO collaboration)}, 
  \bibinfo{journal}{POS (ICHEP 2016)} 
  (\bibinfo{year}{2016}) \bibinfo{pages}{499}. 

\end{thebibliography}

\end{document}